\documentclass[
  reprint,
  amsmath,
  amssymb,
  aps,
  floatfix,
]{revtex4-2}

\usepackage{graphicx}    
\usepackage{dcolumn}     
\usepackage{bm}          
\usepackage{braket}      
\usepackage{multirow}    
\usepackage{makecell}    
\usepackage{booktabs}    
\usepackage{xcolor}      
\usepackage{soul}        
\usepackage{hyperref}    
\hypersetup{
  colorlinks=true,
  allcolors=blue,
}

\soulregister{\cite}{7}  
\soulregister{\eqref}{7} 

\newcommand{\uprightlongdash}[1]{\mathrm{#1}^{\text{--}}}

\newcommand{\tightdots}{.\kern-0.5em.\kern-0.5em.}

\begin{document}

\preprint{APS/123-QED}

\title{Theory of optical spin polarization of axial divacancy and nitrogen-vacancy defects in 4H-SiC}

\author{Guodong Bian}
\author{Gerg\H{o} Thiering}
 \affiliation{HUN-REN Wigner Research Centre for Physics, Institute for Solid State Physics and Optics, PO Box 49, H-1525, Budapest, Hungary} 

\author{\'Ad\'am Gali}
\email{gali.adam@wigner.hun-ren.hu}
 \affiliation{HUN-REN Wigner Research Centre for Physics, Institute for Solid State Physics and Optics, PO Box 49, H-1525, Budapest, Hungary} 
\affiliation{Department of Atomic Physics, Institute of Physics, Budapest University of Technology and Economics, M\H{u}egyetem rakpart 3., 1111 Budapest, Hungary}

\date{\today}
\begin{abstract}
The neutral divacancy and the negatively charged nitrogen-vacancy defects in 4H-silicon carbide (SiC) are two of the most prominent candidates for functioning as room-temperature quantum bits (qubits) with telecommunication-wavelength emission. Nonetheless, the pivotal role of electron-phonon coupling in the spin polarization loop is still unrevealed. In this work, we theoretically investigate the microscopic magneto-optical properties and spin-dependent optical loops utilizing the first-principles calculations. First, we quantitatively demonstrate the electronic level structure, assisted by symmetry analysis. Moreover, the fine interactions, including spin-orbit coupling and spin-spin interaction, are fully characterized to provide versatile qubit functional parameters. Subsequently, we explore the electron-phonon coupling, encompassing dynamics- and pseudo-Jahn--Teller effects in the intersystem crossing transition. In addition, we analyze the photoluminescence PL lifetime based on the major transition rates in the optical spin polarization loop. We compare two promising qubits with similar electronic properties, but their respective rates differ substantially. Finally, we detail the threshold of ODMR contrast for further optimization of the qubit operation. This work not only reveals the mechanism underlying the optical spin polarization but also proposes productive avenues for optimizing quantum information processing tasks based on the ODMR protocol.

\end{abstract}


\maketitle


\section{\label{sec:level1}Introduction}
\begin{figure} 
\includegraphics[width=\linewidth]{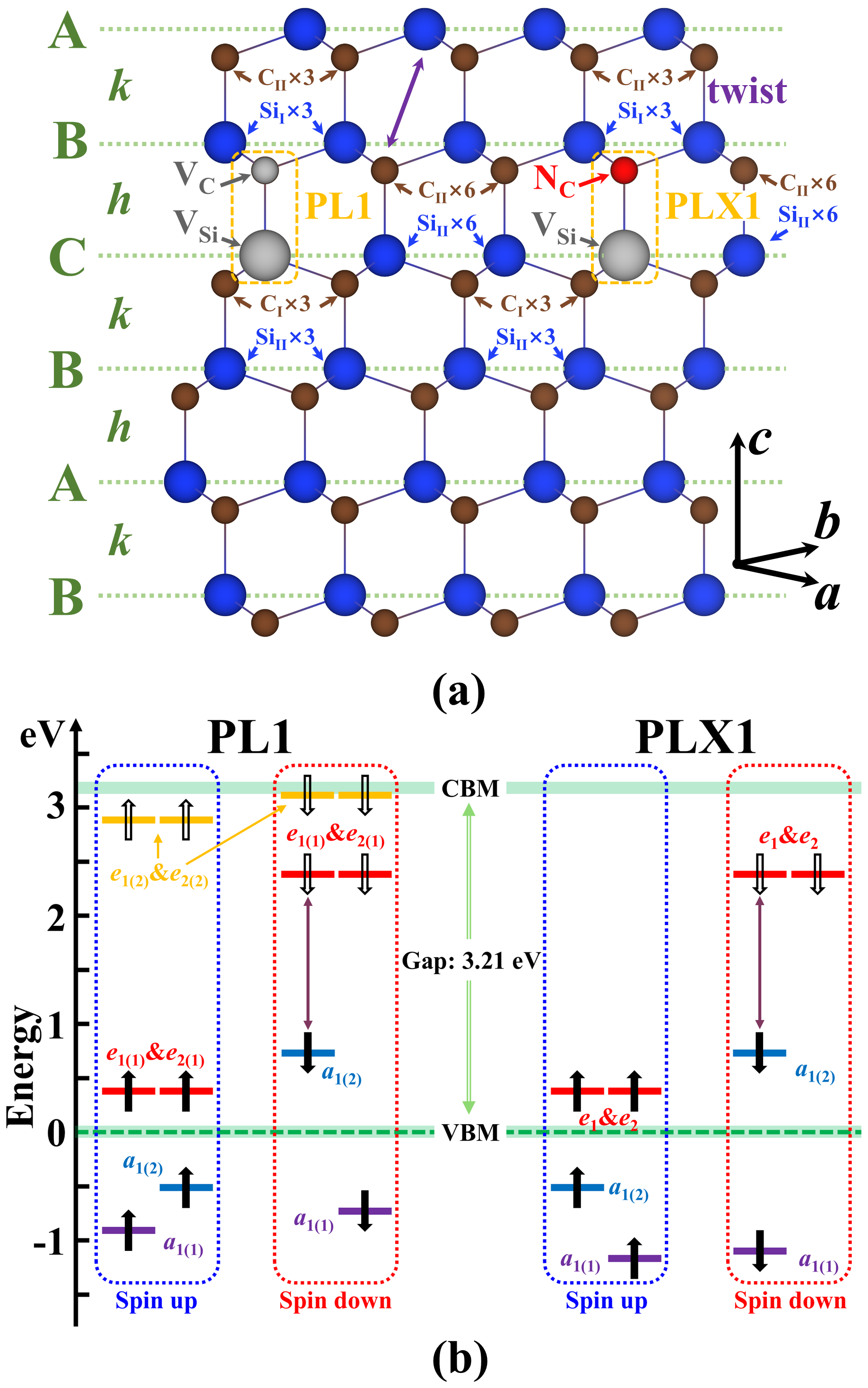}
\caption{\label{fig:geom_KS}(a) Geometry of the PL1 (left) and PLX1 (right) centers. The purple double arrow indicates a 60-degree twist. The arrow combination in the lower right corner is the lattice coordinate system, and $a$, $b$, and $c$ represent the crystal axis. The Si and C with subscripts "I" and "II" are used to label the hyperfine interactions, which will be discussed in the following sections. (b) Hierarchy sketch of the electronic structures for PL1 (left) and PLX1 (right) centers. The horizontal short line is the KS energy level; the solid arrows on it represent occupied electrons, and the hollow arrows represent holes. The gap is the calculated energy band gap of 4H-SiC. CBM (VBM) is the conduction band minimum (valence band maximum). The labels of the levels can be found in the main text. The double arrow between $a_1$ and $e$ levels represents the nature of the transition between ground and excited states.}
\end{figure}

Optically addressable defect spins in solids have attracted significant research interest serving as promising quantum bit (qubit) candidates for emerging quantum information science in the coming noisy intermediate-scale quantum (NISQ) era~\cite{Preskill2018quantumcomputingin,gali2023recent,chatterjee2021semiconductor,wolfowicz2021quantum,atature2018material}. Except for the negatively charged nitrogen-vacancy center in diamond (NV-diamond) well-understood both in theory~\cite{gali2019ab,doherty2011negatively,doherty2013nitrogen,maze2011properties} and experiments~\cite{du2024single,barry2020sensitivity,pezzagna2021quantum,degen2017quantum}, defective silicon carbide (SiC) systems, especially the 4H-SiC polytype~\cite{koehl2011room,csore2022fluorescence,carlos2006annealing,magnusson2005optical,son2006divacancy,zargaleh2016evidence,zargaleh2018electron,li2022room,falk2013polytype,wang2020coherent,crook2020purcell,wolfowicz2017optical,christle2017isolated,christle2015isolated,falk2014electrically,son2022modified,wolfowicz2018electrometry,zhu2021theoretical,mu2020coherent,jiang2023quantum,wang2020experimental,khazen2023nv,csore2017characterization,von2016nv,von2015identification,wang2021optical,anderson2019electrical,ecker2024quantum,csore2021point,anderson2022five}, have attracted an ever growing attention with leveraging an advanced artificial growth and microfabrication techniques of the host SiC crystal. The 4H-SiC is one of the most common polytypes of SiC crystals, with cubic ($k$) and hexagonal ($h$) Si-C bilayers that are stacked by repeating the pattern "ABCB" [see Fig.~\ref{fig:geom_KS}(a)]. Hence, for the neutral divacancy $\mathrm{V_{Si}V_C^0}$ (abbreviated as $\mathrm{VV^0}$) configuration, there are totally four distinct forms: two axial configurations ($hh$, $kk$) and two basal configurations ($hk$ and $kh$). $hh$ and $kk$ configurations have high $C_\text{3v}$ symmetry and are named PL1 and PL2 centers, respectively~\cite{koehl2011room}. The other two possess the $C_\text{1h}$ symmetry and are labeled as PL3 and PL4 centers~\cite{csore2022fluorescence}. All four configurations are named after the four photoluminescence (PL) peaks of UD-2~\cite{carlos2006annealing,magnusson2005optical} that have $S=1$ ground state spin~\cite{son2006divacancy} with zero-phonon line (ZPL) emission at 1132, 1131, 1108, and 1078 nm~\cite{falk2013polytype} for PL1 to PL4 center, respectively. The combination of substitutional nitrogen ($\mathrm{N_{C}}$) and adjacent Si-vacancy ($\mathrm{V_{Si}}$), i.e., the $\mathrm{N_{C}} \mathrm{V_{Si}}$ center (abbreviated as NV center), also has four configurations. After capturing an electron from the crystalline environment, the $\uprightlongdash{NV}$ center in 4H-SiC is formed, of which ZPL peaks yield at 1241, 1242, 1223, and 1180~nm~\cite{zargaleh2018electron} that are labeled by PLX1 to PLX4 similar to the labels of divacancy color centers in 4H-SiC. The two $hh$-axial centers among the four configurations ($hh$, $kk$, $hk$ and $kh$), i.e., the PL1 and PLX1 centers [see Fig.~\ref{fig:geom_KS}(a)], are often favored for their potential in implementing quantum information processing applications and leveraging advantages such as coherent control of spins persist up to elevated temperatures, even room temperature~\cite{koehl2011room,falk2013polytype,li2022room,wang2020coherent} and fluorescence emission around the telecommunication wavelengths~\cite{zargaleh2016evidence,zargaleh2018electron}. Although they have been extensively investigated experimentally~\cite{li2022room,crook2020purcell,magnusson2018excitation,wolfowicz2017optical,christle2017isolated,christle2015isolated,falk2013polytype,falk2014electrically,koehl2011room,shafizadeh2024selection,carlos2006annealing,wang2020coherent,wang2020experimental,zhu2021theoretical,mu2020coherent,zargaleh2016evidence,von2016nv,von2015identification,jiang2023quantum,murzakhanov202314n} and theoretically~\cite{csore2022fluorescence,csore2022photoluminescence,zhu2021theoretical,csore2017characterization,murzakhanov202314n}, the mechanism underlying the optical spin polarization has remained elusive, posing a significant obstacle in implementing quantum information tasks based on the two centers.

In this work, we first present the electronic structure and the resulting multiple basis wavefunctions, which are fundamental to the entire research. Subsequently, we investigate the electronic interaction, encompassing zero-field splitting (ZFS) and spin-orbit coupling (SOC), to elucidate the fine electronic structure. Moreover, we demonstrate that spin-conserving direct transitions involve radiative and direct nonradiative processes between excited and ground-state triplets. We conduct a microscopic examination of the nonradiative spin-dependent intersystem crossing (ISC) transition between states with different spin-multiplicities. Based on the resulting parameters, we assemble a spin polarization optical loop with five key energy levels and the major transitions that occur between them and investigate the optimal optically detected magnetic resonance (ODMR) contrast.

\section{\label{sec:method}Methodology}
We employ the Vienna \textit{Ab-initio} Simulation Package (VASP 5.4.1) code in the framework of density functional theory (DFT) for implanting all atomistic simulations~\cite{kresse1996efficient,kresse1993ab,kresse1996efficiency,paier2006screened}. The Heyd--Scuzeria--Ernzerhof (HSE) hybrid functional with HSE06 parameters~\cite{heyd2003hybrid,krukau2006influence,deak2010accurate} within the DFT technique is applied to reproduce accurate energy band and related information. The PL1 and PLX1 centers are modeled in a standard 576-atom 4H-SiC supercell ($6 \times 6 \times 2$) with a $\Gamma$-point sampling of the Brillouin-zone. The optimized $a$ and $c$ lattice constants are 18.43~\AA\ and 20.10~\AA, respectively. The cutoff energy is set as 420~eV. The atomic configurations are relaxed with the total energy and force thresholds of $1 \times 10^{-4}$~eV and 0.01~eV/\AA. The excited state $\ket{{}^3E}$ is determined using the $\Delta$SCF method~\cite{gali2009theory}, which involves promoting an electron from the $a_1$ orbital to the unoccupied $e$ orbital in the fundamental band gap, as illustrated in Fig.~\ref{fig:geom_KS}(b). Because of that, the conventional Kohn--Sham (KS) DFT cannot adequately describe the $\ket{{}^1{A_1}}$ singlet state because of the high correlation between the two degenerate $e$ orbitals. In this work, the energy and geometry of $\ket{{}^1 A_1}$ are simulated by spinpolarized singlet occupation of the ${e_x}$ orbital~\cite{thiering2018theory}. The ZFS parameters are calculated by employing the VASP projector-augmented-wave implementation of electron spin-spin interaction~\cite{bodrog2013spin} as implemented by Martijn Marsman. Besides, the calculation of SOC parameters uses a noncollinear approach implemented in VASP~\cite{steiner2016calculation}. It is based on the Perdew--Burke--Ernzerhof (PBE) functional~\cite{perdew1996generalized,steiner2016calculation}, making calculations with varying supercell sizes feasible (see Appendix~\ref{appendix:SOC} for details).

\section{\label{sec:electronic_properties}Electronic properties}
It is imperative to fully characterize the electronic properties of the spin system of the color center in order to develop quantum information processing applications. First, we examine the electronic multiple wavefunctions that belong to variable irreducible representation (IR) spaces to lay the foundation for the entire research. We also obtain the electronic structure through first-principles calculations. Additionally, we investigate two types of fine electronic interactions: zero-field splitting and spin-orbit coupling, which jointly determine the fine structure. ZFS provides important parameters for the ODMR protocol, while SOC induces transitions, particularly facilitating intersystem crossing (ISC) via its parallel and perpendicular components. Besides, the hyperfine interaction parameters from isotopes proximate to the core of the point defect are also discussed.

\subsection{\label{sec:basis_functions}Electronic multiplet wavefunctions}
Firstly, we investigate the electronic multiplet wavefunctions using the projection method in group theory~\cite{maze2011properties, doherty2011negatively, bian2022symmetry}. For the PL1 center, there are six dangling bonds in the divacancy [see Fig.~\ref{fig:geom_KS}(a)], which contributes $s{p^3}$ hybrid atomic orbitals to the initial basis. The initial basis vector is $\{s_{1}, s_{2}, s_{2}, c_{1}, c_{2}, c_{3}\}$, where \( s_i \) and \( c_i \) are atomic orbitals of Si and C atoms, respectively. Under the framework of $C_\text{3v}$ symmetry, the projection method is employed to construct the symmetrical molecular orbitals (MOs) belonging to variable IR spaces (${A_1}$, ${A_2}$, and ${E}$ for $C_\text{3v}$ symmetry)  by linear combinations of the atomic orbitals (LCAOs). Ignoring orbital overlap integrals, all the resulting projected MOs are
\begin{subequations}\label{eq:PL1_single}
    \begin{align}
        a_{1(1)} &= \frac{s_1+s_2+s_3}{\sqrt{3}},  & a_{1(2)} &= \frac{c_1+c_2+c_3}{\sqrt{3}},  \\
        e_{1(1)} &= \frac{2c_1-c_2-c_3}{\sqrt{6}}, & e_{1(2)} &= \frac{2s_1-s_2-s_3}{\sqrt{6}}, \\
        e_{2(1)} &= \frac{c_2-c_3}{\sqrt{2}},      & e_{2(2)} &= \frac{s_2-s_3}{\sqrt{2}},
    \end{align}
\end{subequations}
where $a_{1(1)}$ and $a_{1(2)}$ are non-degenerate and belong to $A_1$ IR; $e_{1(1)}$ and $e_{2(1)}$ are degenerate, and so are $e_{1(2)}$ and $e_{2(2)}$. All $e$ MOs belong to the $E$ IR. $\{a_{1(1)}, a_{1(2)}, e_{1(1)}, e_{2(1)}, e_{1(2)}, e_{2(2)}\}$ is a set of symmetric basis vectors of PL1 center. Similarly, based on the initial basis of the PLX1 center $\{n, c_1, c_2, c_3\}$, the resulting projected MOs of the PLX1 center in variable IR spaces are

\begin{subequations}\label{eq:PLX1_single}
\begin{align}
a_{1(1)} &= n,  & a_{1(2)}& =\frac{c_1+c_2+c_3}{\sqrt{3}},  \\
e_{1}    &= \frac{2c_1-c_2-c_3}{\sqrt{6}}, & e_{2}& =\frac{c_2-c_3}{\sqrt{2}} 
\end{align}
\end{subequations}

where $n$ is the nitrogen atom’s $s{p^3}$ orbital, $c$ is from the same site as in PL1 center. The electron singlet basis of PLX1 center $\{a_{1(1)}, a_{1(2)}, e_1, e_2\}$. Although the LCAO method will not accurately describe the orbitals~\cite{doherty2011negatively}, the DFT-HSE06 calculations show the character of this analysis with accurate energy ordering and contributions of every atomic orbital to highly localized states. Both color centers localize six unpaired electrons, but the sources of the electrons are slightly different. In the PL1 center, the three silicon and three carbon nearest neighbor atoms around the divacancy contribute one electron each. In the PLX1 center, the substitutional N contributes two electrons, the three nearest neighbor carbon atoms contribute one electron each, and one electron is captured from the environment. Combining the projection MOs with the first principles calculation KS levels, the energy level diagram is sketched in Fig.~\ref{fig:geom_KS}(b). The calculated band gap is 3.21~eV, close to the experimental value of 3.23~eV~\cite{levinshtein2001properties}.

From Fig.~\ref{fig:geom_KS}(b), for both two centers, the $a_{1(1)}$ level is deeply submerged in the valence band, which means that it will be very difficult to promote electrons from it to other levels with higher energy. Besides, for the PL1 center, the two unoccupied $e_{1(2)}$ and $e_{2(2)}$ are also difficult to be occupied by electrons from lower levels. Hence, in this work, we focus only on three energy levels close to each other in the band gap: ${a}_{1(2)}$ and $e_x$ ($e_{1(1)}$ of PL1, $e_{1}$ of PLX1), $e_y$ ($e_{2(1)}$ of PL1, $e_2$ of PLX1). From Eq.~\eqref{eq:PL1_single} and Eq.~\eqref{eq:PLX1_single}, all selected levels are contributed primarily by dangling bonds of C atoms proximate to the centers. The basis vector of $\{ a, e_x, e_y \}$ shows the $\{ z, x, y \}$ space properties in the color center coordinate system. Fig.~\ref{fig:geom_KS}(b) also shows the total spin $S = 1$ of both centers. Meanwhile, there are a total of four electrons occupying the $a$, $e_x$, and $e_y$ levels, leaving two holes. The multi-electron picture can be equivalently transformed into a double-hole picture, which will significantly simplify the analysis. Starting from the $\{ a, e_x, e_y \}$ basis within the hole notation, all two-hole orbital basis functions $\ket{\varphi}$ are represented as following Eq.\eqref{eq:3A2} to Eq.\eqref{eq:aa}. 

All the ground states possess ($ee$) configuration within the hole notation. The presentations of triplet $\ket{{}^3A_2}$ is
\begin{equation}
\left.
\begin{array}{c}
\ket{{}^3A_2^+} \\
\ket{{}^3A_2^0} \\
\ket{{}^3A_2^-}
\end{array}
\right\}=\frac{1}{2}(\ket{e_+e_-}-\ket{e_-e_+})\otimes
\left\{
\begin{array}{c}
\sqrt{2}\ket{\uparrow\uparrow} \\
\ket{\uparrow\downarrow}+\ket{\downarrow\uparrow} \\
\sqrt{2}\ket{\downarrow\downarrow}
\end{array}
\right.\text{,}
\label{eq:3A2}
\end{equation}
where the label $\{+, 0, -\}$ means the $\{ \ket{1}, \ket{0}, \ket{-1} \}$ spin sub-states. $\ket{e_{\pm}} = \mp \frac{1}{\sqrt{2}} \left( \ket{e_x} \pm i\ket{e_y} \right)$ is a complex combination of the real orbitals $e_x$ and $e_y$. Besides, in the ($ee$) configuration, there are three other singlet ground states with a spin basis function of $\left(\ket{\uparrow\downarrow} - \ket{\downarrow\uparrow}\right)$, the expressions for double-degenerate $\ket{{}^1E_{\mp}}$ and non-degenerate $\ket{{}^1A_1}$
\begin{equation}
\ket{{}^1E_{\mp}}=
\left.
\begin{array}{l}
\ket{e_+ e_+} \\
\ket{e_- e_-}
\end{array} 
\right\}
\otimes\frac{1}{\sqrt{2}}(\ket{\uparrow\downarrow}-\ket{\downarrow\uparrow})\text{,}
\label{eq:1E_mp}
\end{equation}
\begin{equation}
\ket{{}^1A_1}=\left(\ket{e_+ e_-}+\ket{e_- e_+}\right)\otimes \frac{1}{2}(\ket{\uparrow\downarrow}-\ket{\downarrow\uparrow})\text{.}
\label{eq:1A_1}
\end{equation}
Additionally, there is an equivalent set of multi-electron basis functions based on the $(x,y)$ basis as referenced in Ref.~\onlinecite{doherty2011negatively,maze2011properties}. To lay the foundation for the following discussion of the lower branch ISC transition, we introduce a new set of basis vectors $\{\ket{xx}, \ket{xy}, \ket{yy}\}$ as
\begin{equation} \label{eq:PJT_basis}
\left.
\begin{aligned}
\ket{xx}&=\ket{e_x e_x} \\
\ket{xy}&=\frac{1}{\sqrt{2}}(\ket{e_x e_y}+\ket{e_y e_x}) \\
\ket{yy}&=\ket{e_y e_y}
\end{aligned}
\right\}\otimes\frac{1}{\sqrt{2}}(\ket{\uparrow\downarrow}-\ket{\downarrow\uparrow})\text{,}
\end{equation}
to represent the three singlets $\ket{{}^1E_{1,2}}$ and $\ket{^1{A_1}}$ of $(ee)$ configuration as
\begin{align}
\ket{{}^1E_x}&=\frac{1}{\sqrt{2}}(- \ket{xx} + \ket{yy})\text{,} \label{eq:E_x_NEWBASIS} \\
\ket{{}^1E_y}&=\ket{xy}\text{,} \label{eq:E_y_NEWBASIS} \\
\ket{{}^1A_1}&=\frac{1}{\sqrt{2}}(\ket{xx}+\ket{yy})\text{.} \label{eq:1A1_NEWBASIS}
\end{align}

\begin{figure} 
\includegraphics[width=\linewidth]{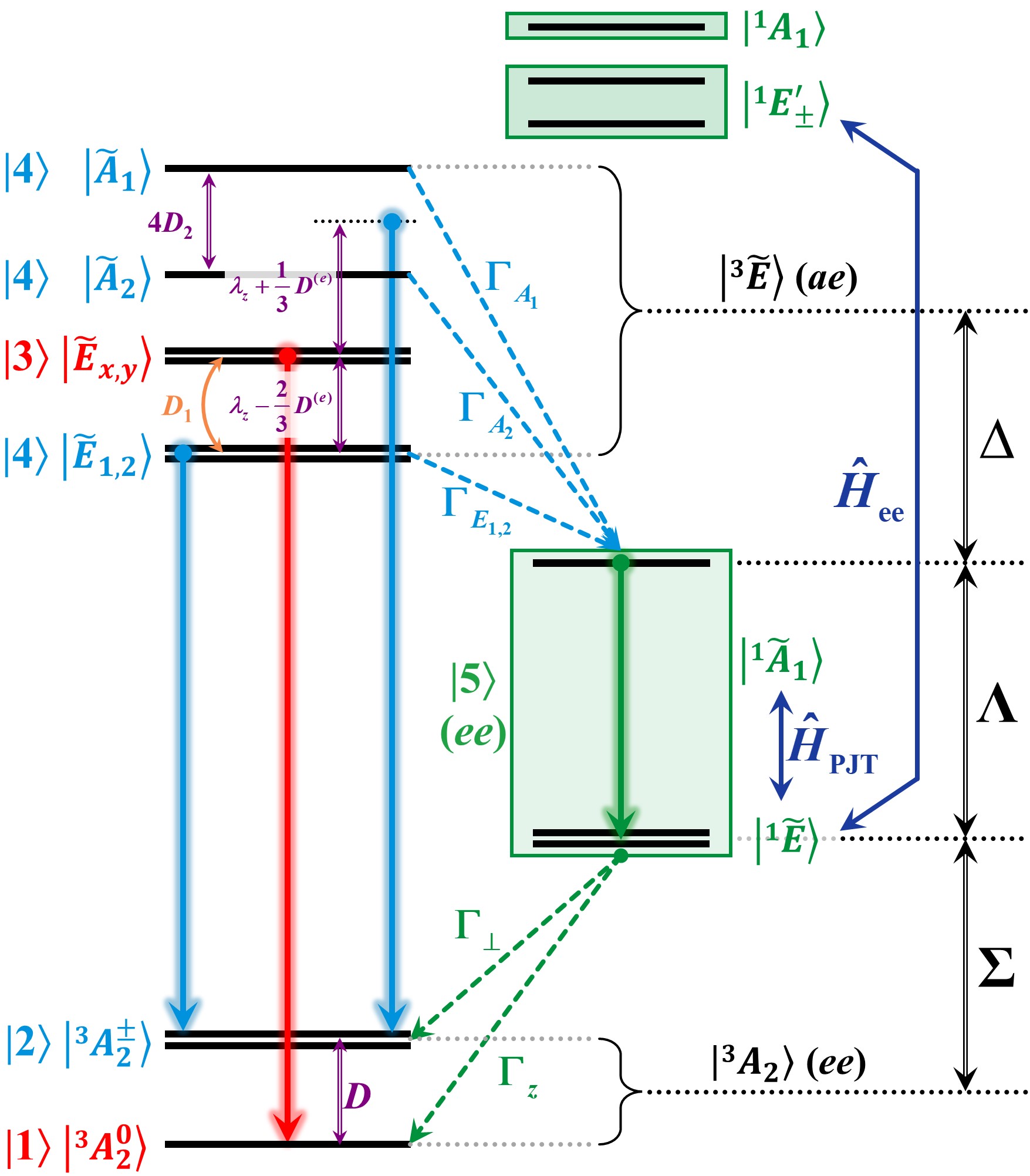}
\caption{\label{fig:fifteen_levels} Hierarchy of multi-electron levels in ascending order shared by PL1 and PLX1 centers. The black horizontal lines represent energy levels. $\lambda_z$ and $D$ represent the axial SOC and ZFS parameters, respectively. The positions of $A_1$ and $A_2$ may be swapped due to the different signs of $D_2$. The glow and dotted arrow lines, respectively, mean radiative and nonradiative transitions. $\Delta$, $\Lambda$, and $\Sigma$ are the energy differences between $\ket{{}^3 \widetilde{E}}$ and $\ket{{}^1 \widetilde{A}_1}$, between $\ket{{}^1 \widetilde{A}_1}$ and $\ket{{}^1 \widetilde{E}}$, and between $\ket{{}^1 \widetilde{E}}$ and $\ket{{}^3A_2}$, respectively. Tilde indicates vibronic state. The dark blue line with double arrows indicates electronic interactions. The orange double arrow curves represent mixing interaction arising from spin-spin interaction.
It is important to note that the energy gap among sublevels of $\ket{{}^3E}$ is not to scale to display fine splitting. In reality, fine splittings on the order of GHz can be disregarded in comparison to $\Delta$, $\Lambda$, and $\Sigma$, which usually have values in the hundreds of meV or even larger.}
\end{figure}

When promoting one electron of spin-minority channel from the occupied in-gap $a$ level to the unfilled in-gap $e$ level, the electronic configuration becomes ($ae$), which corresponds to the first excited states with basis functions of $\ket{{}^3E_\pm}$ and $\ket{E'_\pm}$
\begin{equation} \label{eq:3E}
\ket{{}^3{E_\pm}} = \left.
\begin{array}{c}
\ket{e_+ a} - \ket{a e_+} \\
\ket{e_- a} - \ket{a e_-}
\end{array} 
\right\} \otimes \frac{1}{2} \left\{ 
\begin{array}{c}
\sqrt{2}\ket{\uparrow \uparrow}  \\
\ket{\uparrow \downarrow} + \ket{\downarrow \uparrow} \\
\sqrt{2}\ket{\downarrow \downarrow}
\end{array} \right.\text{,}
\end{equation}
\begin{equation} \label{eq:1E}
\ket{{}^1{E'_\pm}} = 
\left.
\begin{array}{c}
\ket{e_+ a} + \ket{a e_+} \\
\ket{e_- a} + \ket{a e_-}
\end{array}
\right\} 
\otimes \frac{1}{2} (\ket{\uparrow\downarrow} - \ket{\downarrow\uparrow})\text{.}
\end{equation}
$\ket{{}^3E_{\pm}}$ is a multiple degenerate state, which can be divided further into $\ket{E_{1,2}}, \ket{E_{x,y}}$, and $\ket{A_{1,2}}$ states [see Fig.~\ref{fig:fifteen_levels}]. $\ket{E'_\pm}$ is a double-degenerate singlet state. All states of the first excited states are Jahn--Teller (JT) unstable because of the unequal occupation of electrons in the two degenerate $e$ orbitals. When promoting two electrons from $a$ to $e$ orbital, the $a$ orbital will be empty, and two $e$ orbitals will be fully occupied. This is the second excited state with the highest energy and is expressed as
\begin{equation} \label{eq:aa}
\ket{{}^1A'_1} = \ket{aa} \otimes \frac{1}{\sqrt{2}}\left( \ket{\uparrow\downarrow} - \ket{\downarrow\uparrow} \right)\text{.}
\end{equation}

\subsection{\label{sec:SOC_and_ZFS}SOC and ZFS}
The SOC can be divided into axial and transverse components because of the possessed $C_\text{3v}$ symmetry. The axial component dominates the fine modification of the energy levels, especially in the excited states~\cite{doherty2011negatively, doherty2013nitrogen, maze2011properties}, which can be determined by the photoluminescence excitation (PLE) measurement at low temperatures~\cite{christle2017isolated}. The transverse component induces spin-dependent ISC between states with different spin multiplicities, where ISC is one of the most critical prerequisites in spin-dependent fluorescence dynamics. For the two centers, the SOC Hamiltonian $\hat{H}_{\text{SOC}}$ can be written in terms of the angular momentum operators ${\hat l_j}$ by selecting basis of $\{ {{e_+},{e_-},a} \}$ defined in Section~\ref{sec:basis_functions} is
\begin{equation} \label{eq:raising_lowering_SOC_Hamiltonian}
\hat{H}_{\text{SOC}}=\sum_j\left[\frac{1}{2}\lambda_{\bot}(\hat{l}_j^+\hat{s}_j^-+\hat{l}_j^-\hat{s}_j^+)+\lambda_z\hat{l}_j^z\hat{s}_j^z\right]\text{,}
\end{equation}
where ${\lambda_z}$ and $\lambda_{\bot}$ are the respective axial and transverse non-zero matrix elements of the orbital operator $\hat O_k^j$; $\hat{l}_z\ket{e_{\pm}}=\pm\ket{e_{\pm}}$ and $\hat{l}_{\pm}=\hat{l}_x\pm i\hat{l}_y$ are orbital raising and lowering operators with operational relationship of $\displaystyle{\hat{l}_{\pm}\ket{a}=\mp\frac{1}{\sqrt{2}}\ket{e_{\pm}}}$ and $\hat{l}_z\ket{e_{\pm}} = \pm \ket{e_{\pm}}$; ${\hat s_ \pm } = {\hat s_x} \pm i{\hat s_y}$ are spin raising and lower operators. Eq.~\eqref{eq:raising_lowering_SOC_Hamiltonian} can be further rewritten as~\cite{maze2011properties,thiering2018theory}
\begin{equation} \label{eq:Dirc_SOC_Hamiltonian}
\begin{aligned}
\hat{H}_{\text{SOC}}&=\lambda_z(\ket{A_1}\bra{A_1}+\ket{A_2}\bra{A_2}-\ket{E_1}\bra{E_1}-\ket{E_2}\bra{E_2})\\
&\quad+\lambda_{\bot}(\ket{^1E'_+}\bra{^3A_2^+}+\ket{^1E'_-}\bra{^3A_2^-})\\
&\quad+\text{c.c.}
\end{aligned}
\end{equation}
where all the Dirac notations arise from Eq.~\eqref{eq:3A2} to Eq.~\eqref{eq:aa}. We note that
\begin{subequations}\label{eq:lambda_z_lambda_bot}
\begin{align}
\lambda_z &= \braket{{}^3E|\hat{H}_{\text{SOC}}|{}^3E} \simeq \frac{1}{2}\braket{e_+|\hat{H}_{0}|e_+}, \label{eq:lambda_z}\\
\lambda_{\bot} &= \braket{{}^3E|\hat{H}_{\text{SOC}}|{}^1A_1} \simeq \frac{1}{\sqrt{2}}\braket{e_+^\downarrow|\hat{H}_{0}|a^\uparrow}\text{,} \label{eq:lambda_bot}
\end{align}
\end{subequations}
where $H_{\text{SOC}} = \sum_i H_{0i}$, up and down arrows represent the respective spin states and $\ket{e_{\pm}}$ was defined in Eq.~\eqref{eq:3A2} to Eq.~\eqref{eq:aa} and depicted in Fig.~\ref{fig:fifteen_levels}. 

The $\lambda_z$ and $\lambda_{\bot}$ can be obtained by employing the noncollinear magnetic calculations. The quantization axis was set to the ${C_3}$ axis, i.e., the $c$-axis of the 4H-SiC crystal. The geometry comes from spinpolarized DFT calculations possessing high symmetry by smeared electrons in the degenerate $e$ levels, which is fixed when performing the noncollinear calculations because SOC is a tiny perturbation to the system~\cite{thiering2017ab}. The procedure of SOC calculation is under $\Gamma$-point sampling of the Brillouin zone because other $k$ points will introduce ambiguity by reducing the symmetry of orbitals. The strength of ${\lambda_z}$ can be found by comparing the energy difference between electrons occupying the ${e_+}$ and ${e_-}$ levels. This difference is also equal to the splitting of the two double-degenerate $e$ levels when both levels are half occupied. Besides, we perform a scaling method to obtain ${\lambda_z}$ from various sizes of supercells, and then the fitting result ${\lambda _{z0}}$ will belong to an isolated qubit (see Appendix~\ref{appendix:SOC} for detail). The calculated ${\lambda _{z0}}$ for PL1 center is 18.5~GHz, 5.2 times larger than the experimental value of $3.538\pm0.052$~GHZ~\cite{christle2017isolated} measured at 8~K by PLE. The difference between calculated and experimental results can be attributed to the dynamic-JT (DJT) effect, which is beyond the Born--Oppenheimer approximation and reduces the theoretical value by the $p$ Ham reduction factor (abbreviated as $p$ factor) for correcting the $\lambda _{z0}$ result to $\lambda _z = p \cdot \lambda_{z0}$~\cite{ham1968effect,bersuker2006thejahn,bersuker2012vibronic,thiering2017ab}. After being reduced by the $p$ factor, the final ${\lambda _{z0}}$ result for the PL1 center is 1.302~GHz. Besides, the calculated ${\lambda _{z0}}$ of PLX1 is 9.7~GHz and is reduced to 0.85~GHz with $p$ factor (see Appendix~\ref{appendix:SOC} for details of $p$ factor). Though the ${\lambda_\bot}$ can be obtained from the off-diagonal terms of SOC matrix, the calculated value of the ${\lambda_\bot}$ is always much larger in our experience~\cite{thiering2017ab} where the origin of this effect has not yet been identified. Since ${\lambda_\bot}$ remains fixed when simulating the ISC transition process in Section~\ref{sec:ISC_upper}, we utilize the relationship $\lambda_\bot = \lambda_z \times 1.2$ to account for the uncertainty of $\lambda_\bot$ resulting from the $C_\text{3v}$ symmetry~\cite{maze2011properties, goldman2015state}.

The ZFS is a fine splitting arising from the electronic spin-spin interaction among two or more uncoupled electrons without any external magnetic field. The ZFS parameters can be determined by conventional electron spin resonance (ESR)~\cite{gruber1997scanning} and will serve as a direct reference for the microwave frequencies used in experimental ODMR implementations. The Hamiltonian of electronic spin-spin interaction ${\hat H_{ss}}$ is
\begin{equation} \label{eq:spin_spin_Hamiltonian}
\hat{H}_\text{ss} = \frac{\mu_0}{4\pi} \frac{g^2 \beta^2}{r^3} \left[ \hat{s}_1 \cdot \hat{s}_2 - \frac{3(\hat{s}_1 \cdot \hat{r})(\hat{s}_2 \cdot \hat{r})}{r^2} \right]\text{,}
\end{equation}
where $\mu _{0}$ is the magnetic constant, $g$ is the electron Land\'{e} factor, $\beta$ is the Bohr magneton, $\hat s$ is the spin momentum operator, and $\hat r$ is the distance between two electrons. The matrix representation form of ${\hat H_{ss}}$ is~\cite{bian2022symmetry,ivady2014pressure}
\begin{equation} \label{eq:matrix_form_of_Hss}
\begin{aligned}
\hat{H}_\text{ss} &= \hat{S}^\mathrm{T} \cdot ({}^{\mathrm{d}}D) \cdot \hat{S} \\
                  &= D_{xx}\hat{S}_{xx}^2 + D_{yy}\hat{S}_{yy}^2 + D_{zz}\hat{S}_{zz}^2 \\
                  &= D\left(\hat{S}_{zz}^2 - \frac{2}{3}\right) + E\left(\hat{S}_{xx}^2 - \hat{S}_{yy}^2\right)\text{,}
\end{aligned}
\end{equation}
where $\hat S$ is the total spin, $^\mathrm{d} D$ is a second-order trace-less tensor, the superscript "d" indicates diagonal, and $D$ and $E$ are the ZFS parameters in the eigenvalue framework of
\begin{equation} \label{eq:D_E_parameters}
\begin{aligned}
D &= \frac{3}{2} {D}_{zz}, \qquad
E &= \frac{{D}_{yy} - {D}_{xx}}{2}\text{.}
\end{aligned}
\end{equation}
$D$ provides vital evidence for identifying color centers and, further, the frequency of microwave manipulation that causes spin flipping. $E$ indicates the axial symmetry and should be zero for perfect $C_\text{3v}$ symmetry. 

However, the final solution from spin-polarized KS DFT methods of $D$-tensor may not be the eigenstate of the spin operator, which will introduce a discrepancy called spin contamination~\cite{gali2023recent,biktagirov2020spin}. When performing the spin-polarized DFT calculation of PL1 and PLX1 centers with a total spin of 1, the $m_s = 0$ spin configuration will also introduce a non-zero contribution $^{\rm{d}}{D_s}$ to the $D$-tensor of $^{\rm{d}}{D_t}$ (i.e., the ${}^{\mathrm{d}}D$ in Eq.~\eqref{eq:matrix_form_of_Hss}). Then the correct $D$-tensor, the $^{\mathrm{d}}\widetilde{D}$ will be brought by~\cite{gali2023recent,biktagirov2020spin}
\begin{equation} \label{eq:modified_D}
{}^{\mathrm{d}}\widetilde{D} = \frac{{}^{\mathrm{d}}D_t - {}^{\mathrm{d}}D_s}{2}\text{.}
\end{equation}
The corrected ZFS parameters $D$ and $E$ will finally yield using Eq.~\eqref{eq:D_E_parameters} by diagonalizing the $^{\mathrm{d}}\widetilde{D}$.
Prior to addressing spin contamination, the calculated $D$ parameter is 1.93 and 1.95~GHz for PL1 and PLX1 centers, respectively. Following successful spin decontamination~\cite{biktagirov2020spin}, the corrected $D$ parameters are 1.43 and 1.41~GHz, which align well with the experimental values of 1.34~GHz (Ref.~\onlinecite{falk2013polytype}) and 1.33~GHz (Ref.~\onlinecite{von2016nv}) for PL1 and PLX1 centers, respectively.

In addition to the ZFS of the ground states, the ZFS among excited triplet state $\ket{^3E}$ is also discussed. There is also energy level splitting caused by spin-spin interaction, the $D^{(e)}$-tensor (the superscript $(e)$ indicates the excited state) between different spin sublevels in $\ket{^3E}$ [see Fig.~\ref{fig:fifteen_levels}], which is more complicated than the ground triplet state $|^3A_2\rangle$. Utilizing the method implemented in Ref.~\onlinecite{thiering2024nuclear}, the calculated absolute values of $\{ D^{(e)}, D_1^{(e)}, D_2^{(e)} \}$ are respectively $\{2245.17, 26.66, 295.73 \}$ and $\{ 1545.31, 49.13, 331.66 \}$ in MHz for PL1 and PLX1 centers. Furthermore, the Ham reduction factor $q$ should be practically utilized to reflect the $(E \otimes e)$ DJT effect~\cite{thiering2024nuclear,evangelou1980ham} on the $D_1^{(e)}$ and $D_2^{(e)}$-tensors (see Appendix~\ref{appendix:p_q} for detail). The finally $\{ D_1^{(e)}, D_2^{(e)} \}$ values after reduced are $\{ 64.96, 239.98 \}$ and $\{ 128.05, -430.68 \}$ in MHz. We do not currently perform spin decontamination here. Aside from the parallel SOC and $D^{(e)}$-tensor, energy splitting among $\ket{^3E}$ can also be influenced by strain and the external magnetic field to determine the final energy spacing. The effects of strain and the external magnetic field are not within the scope of this work and will not be addressed for now.

\subsection{\label{ sec:hyperfine}Hyperfine parameters}
The hyperfine interaction between electron spins of color centers and proximate nuclear spins is investigated to expand diversified QIS applications, for instance, optically pumped dynamic nuclear polarization for potential SiC-based quantum memories \cite{falk2015optical} and linking single photon emitters with nuclear registers via divacancy center\cite{bourassa2020entanglement}.Table~\ref{tab:hyp} displays all the calculated hyperfine parameters for the first and second neighbor isotopes [see Fig.~\ref{fig:geom_KS}(a)] to the color centers. $\text{C}_{\text{\mbox{I}}}\times 3$ means there are three nearest neighbor $^{13}\text{C}$ nuclear spins with equivalent positions. $\text{C}_{\text{\mbox{II}}}$ means the next-nearest $^{13}\text{C}$ nuclear spins. Besides, the $\text{C}_{\text{\mbox{II}}}$ is divided into two categories according to the relative position in the color center: $\text{C}_{\text{\mbox{II}}}\times 6$ and $\text{C}_{\text{\mbox{II}}}\times 3$, which will show different hyperfine interaction strength. The Si atoms with "I" and "II" indicate $^{29}\text{Si}$ nuclear spins and possess the same location information as $^{13}\text{C}$. In Fig.~\ref{fig:geom_KS}(a), due to the screenshot angle, some atoms are not visible. All the calculated hyperfine parameter results of the PL1 center are consistent with data in Ref.~\onlinecite{son2006divacancy,falk2015optical}. The most critical hyperfine parameters of the PLX1 center are for the substitutional nitrogen isotopes. The calculated $^{14}\text{N}$ hyperfine parameters of PLX1 center are shown in Table~\ref{tab:hyp}, which agree with experimental results of ${A_\parallel}$: 1.17~MHz in Ref.~\onlinecite{murzakhanov202314n}, 1.23~MHz in Ref.~\onlinecite{von2016nv,zargaleh2018electron} and around 1.3~MHz in Ref.~\onlinecite{wang2020coherent}, and it is also consistent with theoretical results~\cite{csore2017characterization} well. Besides, we also calculated the $^{15}\text{N}$ hyperfine parameters to provide an additional pathway for QIS application based on the PLX1 center's nuclear spins.
\begin{table}
\caption{\label{tab:hyp}The calculated total hyperfine coupling parameters for PL1 and PLX1 centers with units in megahertz (MHz). See Fig.~\ref{fig:geom_KS}(a) for details on the isotopic subscripts. Notably, hyperfine parameters for nitrogen isotopes occur in the PLX1 center.}
\begin{ruledtabular}
\renewcommand{\arraystretch}{1.2}  
\begin{tabular}{cccc}
\multirow{2}{*}{Sites} & \multicolumn{3}{c}{PL1/PLX1}  \\
\cline{2-4} 
	&	$A_{xx}$				&	$A_{yy}$			&	$A_{zz}$				\\	\hline
$^{14}\text{N}$        & \makebox[2.1cm][r]{/ $-1.67$}  & \makebox[2.1cm][r]{/ $-1.67$}  & \makebox[2.1cm][r]{/ $-1.73$}  \\
$^{15}\text{N}$        & \makebox[2.1cm][r]{/ $-0.54$}  & \makebox[2.1cm][r]{/ $-0.54$}  & \makebox[2.1cm][r]{/ $-0.56$}  \\
$^{29}\text{Si}_{\text{\mbox{I}}}\times 3$	&	0.52 	/	0.54 	&	$-0.48$ /	$-0.01$ &	0.56 	/	0.64 		\\
$^{13}\text{C}_{\text{\mbox{I}}}\times 3$	&	48.89 	/	45.95 	&	48.20 	/	45.23 	&	119.15 	/	117.07 		\\
$^{29}\text{Si}_{\text{\mbox{II}}}\times 6$	&	9.44 	/	9.89 	&	8.07 	/	8.69 	&	10.25 	/	10.73 		\\
$^{29}\text{Si}_{\text{\mbox{II}}}\times 3$	&	11.35 	/	11.76 	&	11.26 	/	11.68 	&	11.69 	/	12.10 		\\
$^{13}\text{C}_{\text{\mbox{II}}}\times 6$	&	0.97 	/	0.44 	&	0.85 	/	0.39 	&	2.03 	/	1.26 		\\
$^{13}\text{C}_{\text{\mbox{II}}}\times 3$	&	$-0.34$ /	$-0.10$ &	$-0.12$ /	$-0.09$ &	\makebox[0.6cm][r]{$-0.35$} /0.15 		\\
\end{tabular}
\end{ruledtabular}
\end{table}

\section{\label{sec:transition_loop}Spin-dependent optical loop}
The ODMR technique is essential for achieving qubit applications by utilizing spin-dependent fluorescence dynamics within a customized optical loop framework, where the major transition rates among key levels are essential. The nonradiative ISC transition dominated by electron-phonon coupling plays a pivotal role in the ODMR protocol for achieving optical spin polarization. Additionally, the ODMR contrast ratio is closely related to both the nonradiative transition and radiative transition rates. In this section, both the upper branch of ISC transitions (between $\ket{^3E}$ and $|^1A_1\rangle$) mediated by the DJT effect and of the lower branch (between $\ket{^1E_{1,2}}$ and $\ket{^3A_2}$) mediated by the joint effect of DJT and PJT are demonstrated separately. Then, the radiative and direct nonradiative transitions from $\ket{^3E}$ to $\ket{^3A_2}$ are also investigated. Finally, by combining all the results with published experimental data, a brief discussion on ODMR contrast is provided.


\subsection{\label{sec:ISC_upper}Upper branch of ISC}

\begin{figure}[h]
\includegraphics[width=\linewidth]{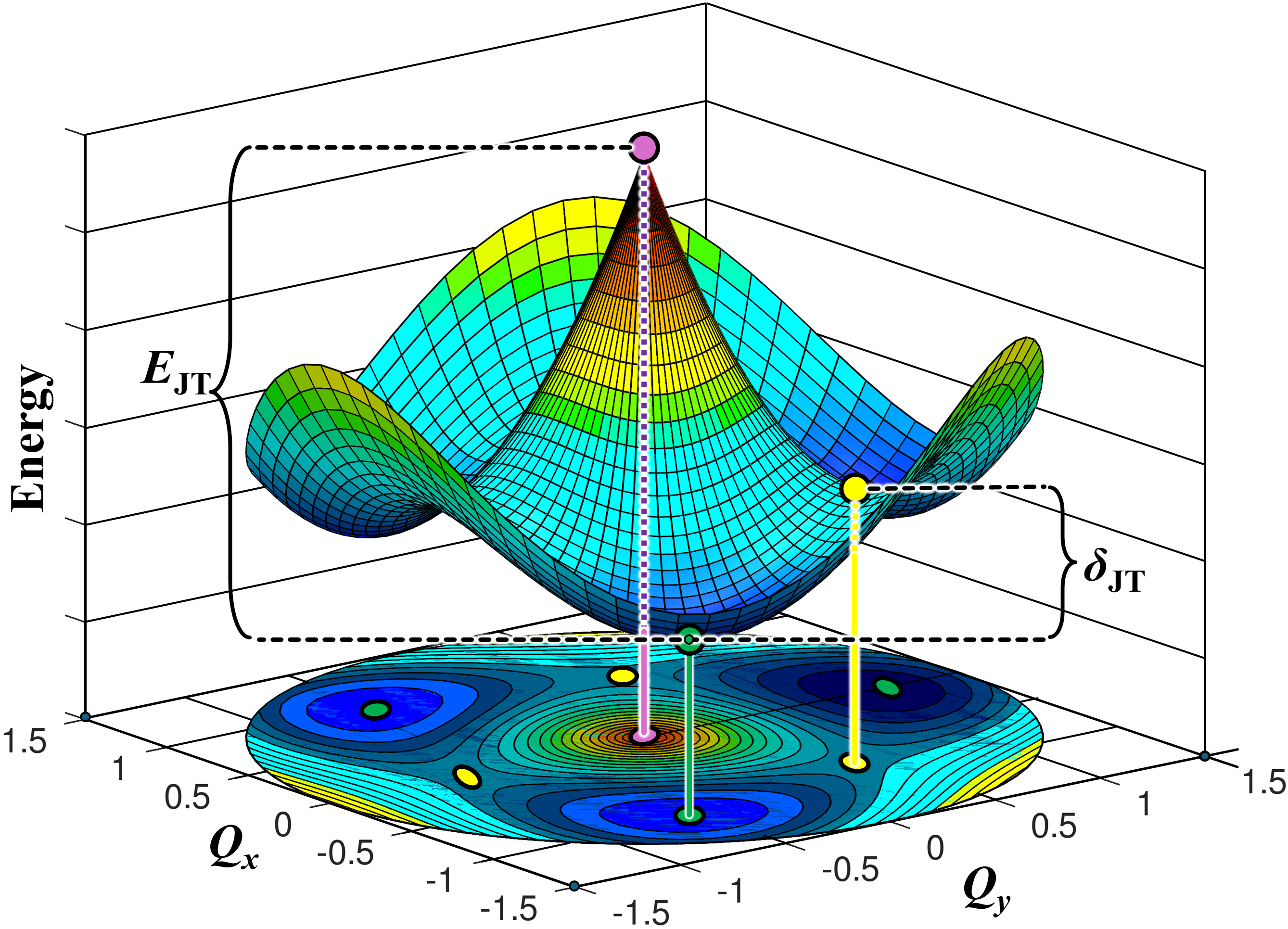}
\caption{\label{fig:APES}APES of the quadratic DJT system for $\ket{^3E}$ of PL1 center. $Q_{x}$ and $Q_{y}$ represent the degenerate $e$ phonons. $E_{\text{JT}}$ is the linear DJT energy, which is the difference between the high ($C_{3\text{v}}$) geometry energy and minima ($C_{1\text{h}}$). ${\delta _{\rm{JT}}}$ is the barrier energy arising from the second-order DJT effect. Green and yellow circles, respectively, indicate the configuration coordinates of three minima and three barrier points.}
\end{figure}

Combined with our analysis of electronic properties in Section~\ref{sec:electronic_properties}, the high symmetry of the orbitally doubly-degenerate ${}^3E$ excited state will be broken when coupling $e$ phonons or quasi-local vibration modes. This is the so-called $(E \otimes e)$ DJT system~\cite{thiering2017ab,bersuker2006thejahn,bersuker2012vibronic}. By introducing two phonon operators $\hat{x}$ and $\hat{y}$, the $(E \otimes e)$ DJT Hamiltonian is~\cite{thiering2017ab}
\begin{align}\label{eq:DJT_Hmiltonian}
\hat{H}_\text{DJT} &= \hbar\omega_e (\hat{a}_x^\dagger \hat{a}_x + \hat{a}_y^\dagger \hat{a}_y + 1) + F (x \hat{\sigma}_z - y \hat{\sigma}_x) \nonumber \\
        &\quad + G [(x^2 - y^2) \hat{\sigma}_z + 2xy \hat{\sigma}_x]\text{,}
\end{align}
where ${\hbar\omega_e}$ is the energy of $e$ mode which will drive the distortion, $F$ and $G$ are linear and second-order electron-vibration
coupling related terms, $\displaystyle{\hat x = \frac{1}{{\sqrt 2}}(a_x^\dag+{a_x}),{\kern 1pt} {\kern 1pt} {\kern 1pt} {\kern 1pt} {\kern 1pt} {\kern 1pt} \hat y = \frac{1}{{\sqrt 2}}(a_y^\dag+{a_y})}$ are the two dimensionless non-Hermitian operators, $\ket{0,0}$, $\ket{1,0}$ and $\ket{0,1}$ are selecting as basis vectors, $\sigma$ is the Pauli matrix. The $F$ and $G$ are directed obtained by
\begin{equation}\label{eq:F_and_G}
F=\sqrt{2\hbar\omega_e E_{\text{JT}}}, \quad G=\frac{\delta_{\text{JT}}\hbar\omega_e}{2E_{\text{JT}}}\text{.}
\end{equation}
The ${\hbar\omega_e}$ is derived directly by parabola fitting in the actual adiabatic potential energy surface (APES) of the quadratic DJT system. All the calculated parameters in Eq.~\eqref{eq:F_and_G} are shown in Table~\ref{tab:DJT}. Additionally, the APES of the PL1 center is plotted for visualization [see Fig.~\ref{fig:APES}]. For simplicity, we only show the $Q$ configuration coordinates of the PL1 center here; in fact, the case of the PLX1 center is very close to the PL1 center. 
\begin{table}[h]
\caption{\label{tab:DJT} The DJT effect parameters of PL1 and PLX1 centers with unit of meV. All results are valid at 0 K.}
\begin{ruledtabular}
\renewcommand{\arraystretch}{1.2}
\begin{tabular}{cccccc}
     & ${E_{\rm{JT}}}$ & $\delta$ & $\hbar\omega_e$ & \textit{F} & \textit{G} \\
\hline 
PL1  &73.62 &18.24  &46.21   &76.43   &3.27  \\
PLX1 &79.22 &23.01  &54.78   &84.88   &4.65  \\
\end{tabular}
\end{ruledtabular}
\end{table}

The electron-phonon coupling coefficients $c_{nm}$ and $d_{nm}$ can be obtained by solving Eq.~\eqref{eq:DJT_Hmiltonian} with parameters displayed in Table~\ref{tab:DJT}. The solution expands as
\begin{equation} \label{eq:DJT_expension}
|\Psi_\pm\rangle=\sum_{nm}\left(c_{nm}|E_\pm\rangle\otimes|n,m\rangle+d_{nm}|E_\mp\rangle\otimes|n,m\rangle\right)\text{.}
\end{equation}
The upper branch of ISC transitions from triplet $\ket{^3E}$ to singlet $|^1A_1\rangle$ mediated by transverse SOC follow the Fermi's golden rule and the transition rate can be expressed as~\cite{thiering2017ab,goldman2015state,goldman2015phonon}
\begin{equation} \label{eq:Gamma_1A1_upper}
\Gamma_{^1A_1} = 4\pi \hbar \lambda_\bot^2 F(\Delta),
\end{equation}
where $F(\Delta)$ is the vibrational overlap function~\cite{goldman2015state}, which is the energy-dependent density of states multiplied by the overlap of the vibrational states between $\ket{^3E}$ and $|^1A_1\rangle$ with energy spacing of $\Delta$. The first order ISC transition occurs only between $|A_1\rangle$ sub-state of $\ket{^3E}$ (see $\ket{4}$ in Fig.~\ref{fig:fifteen_levels}) and $|^1A_1\rangle$ ($\ket{5}$ in Fig.~\ref{fig:fifteen_levels}) described by Eq.~\eqref{eq:Gamma_1A1_upper}, with an assumption of that the ${\lambda _\bot}$ remains fixed independently of the coordinates of the atoms~\cite{thiering2017ab}. Hence, we not only use calculated numerical data of the ${\lambda _\bot}$ described in Section~\ref{sec:SOC_and_ZFS}, but also experimental data available~\cite{christle2017isolated}. For meticulous investigation of the ISC transition between $\ket{^3E}$ and $|^1A_1\rangle$, the nature of $\ket{^3E}$ invoking DJT should be involved, which will bring to a second order of the ISC transitions. The four wavefunctions of electron-phonon coupled triple excited states with magnetic quantum number ${m_s} = \pm 1$ in the Born--Oppenheimer basis of symmetry-adapted terms $\left\{ |\widetilde{A}_1\rangle, |\widetilde{A}_2\rangle, |\widetilde{E}_1\rangle, |\widetilde{E}_2\rangle \right\}$ are variants of Eq.~\eqref{eq:DJT_expension} and take the forms below
\begin{subequations}\label{eq:ISC_upper}
    \begin{align}
        \ket{\widetilde{A}_1} &= \frac{1}{\sqrt{2}}\left(\ket{\Psi_-} \otimes \ket{\uparrow\uparrow} - \ket{\Psi_+} \otimes \ket{\downarrow\downarrow}\right) \nonumber \\
        &= \sum_i \bigg[c_i \ket{A_1} \ket{\varrho_i(A_1)} + f_i \ket{A_2} \ket{\varrho_i(A_2)} \nonumber \\
        &\quad + \frac{d_i}{\sqrt{2}} \left(\ket{E_1} \ket{\varrho_i(E_1)} + \ket{E_2} \ket{\varrho_i(E_2)}\right)\bigg]\text{,} \label{eq:ISC_upper_A1}
    \end{align}
    \begin{align}
        \ket{\widetilde{E}_1} &= \frac{1}{\sqrt{2}}\left(\ket{\Psi_-} \otimes \ket{\downarrow\downarrow} - \ket{\Psi_+} \otimes \ket{\uparrow\uparrow}\right) \nonumber \\
        &= \sum_i \bigg[c_i \ket{E_1} \ket{\varrho_i(A_1)} + f_i \ket{E_2} \ket{\varrho_i(A_2)} \nonumber \\
        &\quad + \frac{d_i}{\sqrt{2}} \left(\ket{A_1} \ket{\varrho_i(E_1)} + \ket{A_2} \ket{\varrho_i(E_2)}\right)\bigg]\text{,} \label{eq:ISC_upper_E1}
    \end{align}
    \begin{align}
        \ket{\widetilde{E}_2} &= \frac{1}{\sqrt{2}}\left(\ket{\Psi_-} \otimes \ket{\downarrow\downarrow} + \ket{\Psi_+} \otimes \ket{\uparrow\uparrow}\right) \nonumber \\
        &= \sum_i \bigg[c_i \ket{E_2} \ket{\varrho_i(A_1)} + f_i \ket{E_1} \ket{\varrho_i(A_2)} \nonumber \\
        &\quad + \frac{d_i}{\sqrt{2}} \left(\ket{A_1} \ket{\varrho_i(E_2)} + \ket{A_2} \ket{\varrho_i(E_1)}\right)\bigg]\text{,} \label{eq:ISC_upper_E2}
    \end{align}
    \begin{align}
        \ket{\widetilde{A}_2} &= \frac{1}{\sqrt{2}}\left(\ket{\Psi_-} \otimes \ket{\uparrow\uparrow} + \ket{\Psi_+} \otimes \ket{\downarrow\downarrow}\right) \nonumber \\
        &= \sum_i \bigg[c_i \ket{A_2} \ket{\varrho_i(A_1)} + f_i \ket{A_1} \ket{\varrho_i(A_2)} \nonumber \\
        &\quad + \frac{d_i}{\sqrt{2}} \left(\ket{E_1} \ket{\varrho_i(E_1)} - \ket{E_2} \ket{\varrho_i(E_2)}\right)\bigg]\text{,} \label{eq:ISC_upper_A2}
    \end{align}
\end{subequations}
where the calculated ${c_i}$, ${d_i}$, ${f_i}$ values are taken from Table~\ref{tab:all_coeff} in Appendix~\ref{appendix:p_q}, and the expressions of symmetry-adapted vibrational wavefunctions $|\varrho_i(\Gamma_i)\rangle$ are shown in Table~\ref{tab:upper_phonon}. Furthermore, by taking account the DJT effect in $\ket{^3E}$, the degenerate $|\widetilde{A}_2\rangle, |\widetilde{E}_1\rangle, |\widetilde{E}_2\rangle$ vibronic wavefunctions containing $|A_1\rangle$ could induce the second order ISC transitions to $|^1A_1\rangle$ with expressions as
\begin{subequations}\label{eq:ISC_rates_upper}
\begin{align}
\Gamma_{A_1} &= 4\pi\hbar\lambda_\bot^2\sum_{i=1}^\infty\left[c_i^2F\left(\Delta - n_i\hbar\omega_e\right)\right]\text{,} \label{eq:GammaA1}\\
\Gamma_{E_{1,2}} &= 4\pi\hbar\lambda_\bot^2\sum_{i=1}^\infty\left[\frac{d_i^2}{2}F\left(\Delta - n_i\hbar\omega_e\right)\right]\text{,} \label{eq:GammaE12}\\
\Gamma_{A_2} &= 4\pi\hbar\lambda_\bot^2\sum_{i=1}^\infty\left[f_i^2F\left(\Delta - n_i\hbar\omega_e\right)\right]\text{.} \label{eq:GammaA2}
\end{align}
\end{subequations}

The calculation of the ISC rates based on Eq.~\eqref{eq:ISC_rates_upper} needs a determination of the unknown $\Delta$, so it was set as a parameter in the following analysis. However, the current HSE06-DFT method cannot explicitly simulate the $|^1A_1\rangle$. The energy and geometry of the $|^1A_1\rangle$ are roughly approximated by the non-spinpolarized DFT calculations of closed-shell $|xx\rangle$ in Eq.~\eqref{eq:PJT_basis}.The feasibility of this method has been verified in Refs.~\onlinecite{thiering2017ab, thiering2018theory}. The overlap function $F(\Delta)$ is approximated from the phonon sideband in the PL spectrum within the Huang--Rhys approximation of the Franck--Condon theory (see Supplementary Material of Ref.~\onlinecite{thiering2017ab} in detail). The original PL spectrum is in the form of $\omega^3 S(\omega)$, and we used $F(\omega) = S(\omega)$ instead. Under this assumption, there are only $a_1$ phonons considered in the ISC process, where the contribution of $e$ phonons is responsible for the DJT nature of the $\ket{^3E}$ and work in the form of ${c_i}$, ${d_i}$, ${f_i}$ coefficients as shown in Eq.~\eqref{eq:ISC_rates_upper}. Hence, for calculating $F(\omega)$, we prefer a high symmetry geometry without any DJT feature of $\ket{^3E}$ by the smeared occupation of electrons in the $e$ levels. 

\begin{figure}[h]
\includegraphics[width=\linewidth]{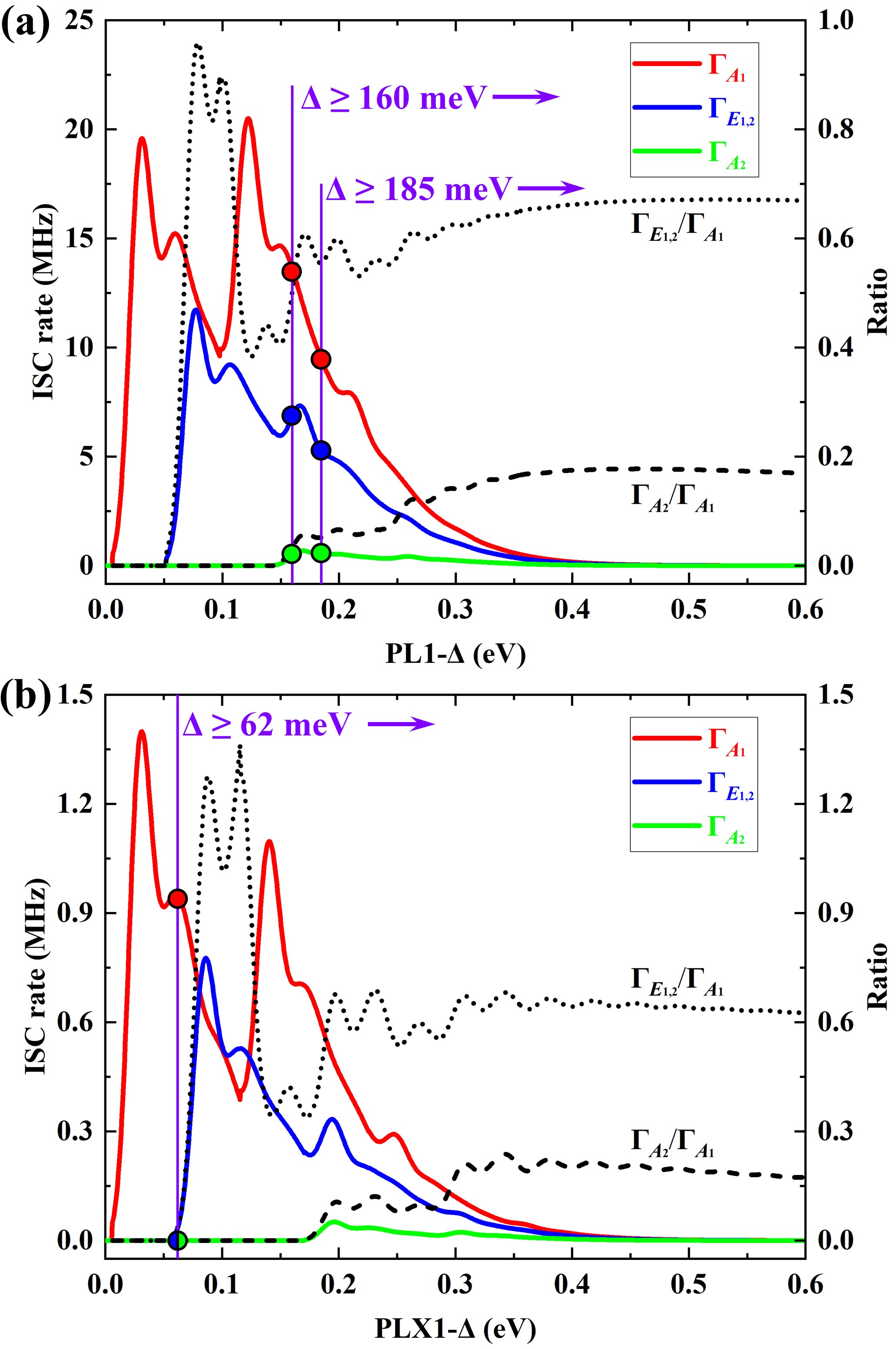}
\caption{\label{fig:ratio_upper}The upper branch of ISC transition rates $\Gamma$ and ratios for (a) PL1 center and (b) PLX1 center. The vertical purple lines correspond to $\Delta$ (see Fig.~\ref{fig:fifteen_levels}) values. For the PL1 center, the $\Delta$ = 160~meV from DFT calculation and 185~meV from Ref.~\onlinecite{bockstedte2018ab}. For the PLX1 center, the $\Delta$ = 62~meV comes from the DFT calculation. The colored circles represent the values of ISC rates at specific $\Delta$.}
\end{figure}

The upper branch of ISC transition originates from different states of $\ket{^3E}$, and it is imperative to provide all rates and ratios between them in relation to the gap energy $\Delta$. All the calculated ISC rates are depicted in Fig.~\ref{fig:ratio_upper}. We observed that although the DJT nature was invoked in triplet excited states, the contribution of ISC $\Gamma_{A_2}$ remains smaller compared to $\Gamma_{A_1}$ and $\Gamma_{E_{1,2}}$ in both centers due to the smaller value of $\sum f_i^2$. For the PL1 center, $\Delta$ = 160~meV is obtained from this DFT calculation, while an additional $\Delta$ = 185~meV comes from the multiconfigurational DFT approach~\cite{bockstedte2018ab}. For $\Delta$ = 160~meV, we found that $\Gamma_{A_1}$ = 13.60~MHz and $\Gamma_{E_{1,2}}$ = 6.85~MHz, with ratio of $\Gamma_{E_{1,2}} / \Gamma_{A_1}$ = 0.50; where for $\Delta$ = 185~meV, $\Gamma_{A_1}$ = 9.46~MHz and $\Gamma_{E_{1,2}}$ = 5.25~MHz, with ratio of $\Gamma_{E_{1,2}} / \Gamma_{A_1}$ = 0.55. The calculated rates of the PL1 center consistently show the reported effective dark state time of 60.7~ns at 5 K~\cite{crook2020purcell}. Moreover, Ref.~\onlinecite{li2022room} reported a mixed transition rate of approximately 14~MHz at room temperature, suggesting that the influence of temperature on ISC rates is relatively insignificant. As for PLX1 center at $\Delta$ = 62~meV, $\Gamma_{A_1}$ is 0.95~MHz, $\Gamma_{E_{1,2}}$ is 0.03 MHz, and $\Gamma_{A_2}$ is almost zero, which shows same order of magnitude to experimental results in Refs.~\onlinecite{wang2020coherent,wang2020experimental}. $\Delta$ = 62~meV is smaller than $\Delta$ = 160~meV of PL1 center and maybe because there are more components from $a$ states than $e_x$ state contributing to $\ket{xx}$, where $a$ has higher energy than $e_x$ within the hole notation. The accuracy and reliability of the DJT parameters in $\ket{^3E}$ are highly credible, and a more precise determination of $\Delta$ values in the future may lead to an even more accurate estimation of the ISC rates.

\begin{table*} 
\caption{\label{tab:upper_phonon}Expressions of symmetry-adapted vibration wave-functions individually defined in Eq.~\eqref{eq:ISC_upper}. $i$ is the sum of quantum numbers of $x$ and $y$ phonons. "/" means no quantum number there. For example, $|\varrho_i(A_2)\rangle$ phonon function starts having a quantum number from $i = 3$. The expressions of the states are not normalized for the sake of brevity.} 
\begin{ruledtabular}
\renewcommand{\arraystretch}{1.2}
\begin{tabular}{ccccc}
 $i = m + n$  & $|\varrho_i(A_1)\rangle$  & $|\varrho_i(E_1)\rangle$  & $|\varrho_i(E_2)\rangle$  & $|\varrho_i(A_2)\rangle$ \\ 
 \hline
 0  & $\ket{0,0}$   & /   & /  & / \\
 1  & /  & $\ket{x,0}$ & $\ket{0,y}$  & /   \\
 2  & $\ket{x^2+y^2}$ & $\ket{x^2-y^2}$  & $\ket{xy+yx}$    & /  \\
 3  & $\ket{x(x^2-3y^2)}$   & $\ket{(x^2+y^2)x}$ & $\ket{(x^2+y^2)y}$ & $\ket{y(3x^2-y^2)}$   \\
 4  & $\ket{(x^2+y^2)^2}$  & $\ket{(x^2+y^2)(x^2-y^2)}$ &  $\ket{(x^2+y^2)(xy+yx)}$   & $\cdots$   \\
 5  & $\cdots$ & $\ket{x^4-6x^2y^2+y^4}$ & $\cdots$  & $\cdots$   \\
 $\cdots$    & $\cdots$  & $\cdots$ & $\cdots$ & $\cdots$ \\
\end{tabular}
\end{ruledtabular}
\end{table*}

\subsection{\label{sec:ISC_lower}Lower branch of ISC}
The lower branch of ISC transition between the double-degenerate $\ket{^1E_{1,2}}$ (abbreviated as $\ket{^1E}$ here and after) and triplet $\ket{^3A_2}$ is more complex than the upper branch. Because $\ket{^1E}$ and $\ket{^1A_1}$ possess different IR spaces, only the symmetry-distorting $e$ vibration modes couple the two states. This is the so-called pseudo-JT (PJT) effect~\cite{bersuker2012vibronic, bersuker2006thejahn}. Employing the basis shown in Eq.~\eqref{eq:PJT_basis}, the expression of Hamiltonian including the electronic component $\hat{H}_\text{e}$, harmonic oscillator component $\hat H_{\text{osc}}$, and PJT component $\hat{H}_{\text{PJT}}$ is
\begin{align}
\label{eq:PJT_H}
\hat{H} &= \hat{H}_\text{e} + \hat{H}_{\text{osc}} + \hat{H}_{\text{PJT}} \nonumber \\
       &= \frac{\Lambda_e}{2}
\begin{bmatrix}
1 & 0 & 1 \\
0 & 0 & 0 \\
1 & 0 & 1
\end{bmatrix}
+ \hbar \omega_E \left(a_x^\dagger a_x + a_y^\dagger a_y + 1\right) \nonumber \\
       &\quad + \widetilde{F} \left(\hat{\sigma}_z \hat{x} - \hat{\sigma}_x \hat{y}\right)\text{,}
\end{align}
where ${\Lambda_e}$ is the energy gap between $\ket{^1E}$ and $\ket{^1A_1}$ when the electron-phonon interaction is not considered, ${\hbar\omega_E}$ is the energy of $e$ mode of PJT, $\hat{x}$ and $\hat{y}$ are dimensionless coordinates and defined in Eq.~\eqref{eq:DJT_Hmiltonian} with frequency of $\omega_E$, $\tilde F$ is the cumulative electron-phonon coupling, $\hat\sigma_z$ and $\hat\sigma_y$ are spin operators of the angular momentum $L = 1$ in the PJT interaction with the following form
\begin{equation}
\label{eq:sigmasPJT}
\hat{\sigma}_z = \begin{bmatrix}
1 & 0 & 0 \\
0 & 0 & 0 \\
0 & 0 & -1
\end{bmatrix}, \quad
\hat{\sigma}_x = \frac{-1}{\sqrt{2}} \begin{bmatrix}
0 & 1 & 0 \\
1 & 0 & 1 \\
0 & 1 & 0
\end{bmatrix}\text{.}
\end{equation}

Besides the PJT interaction in $\ket{^1E}$, there is also a dynamic electron-electron correlation between the $\ket{{}^1E'}$ and $\ket{^1E}$, and the DJT effect will also be involved. The electron-electron correlation happens among two states with the same total symmetry, even themselves. In this work, we mainly focus on the mixture of $\ket{{}^1E'}$ and $\ket{^1E}$, which will allow the $\Gamma_{\bot} = \Gamma_{\pm} + \Gamma_{\mp}$. We introduced a mixing coefficient $C$ for describing the multi-determinant singlet state $\ket{{}^1\bar{E}}$~\cite{thiering2018theory} quantitatively as
\begin{equation}
\ket{{}^1\bar{E}}=C\ket{{}^1E}+\sqrt{1-C^2}\ket{{}^1E'}\text{.}
\label{eq:mix_C}
\end{equation}
Based on Eq.~\eqref{eq:mix_C}, the $\ket{^1E}$ will carries the DJT character by the extent of $(1-C^2)$, which also indicates the contribution of $\ket{{}^1E'}$ in $|{}^1\bar{E}\rangle$. The DJT Hamiltonian of $\ket{{}^1E'}$ is
\begin{equation}
\hat{H}_{\text{DJT}}=F_2(\bar{\sigma}_z\hat{X}-\bar{\sigma}_x\hat{Y})\text{,}
\label{eq:DJT_Hmiltonian_lower_branch}
\end{equation}
where $F_2$ is the electron-phonon coupling of DJT, ${\bar\sigma_z}$ and ${\bar\sigma_x}$ are spin operators of the $L = 1$ angular momentum spinning in the two-dimensional $\ket{{}^1E'}$ space with the form of
\begin{subequations}\label{eq:sigma}
\begin{align}
\bar{\sigma}_z&=\ket{E'_x}\bra{E'_x}-\ket{E'_y}\bra{E'_y}\text{,} \label{eq:sigma_z} \\
\bar{\sigma}_x&=\ket{E'_x}\bra{E'_y}+\ket{E'_y}\bra{E'_x}\text{,} \label{eq:sigma_x}
\end{align}
\end{subequations}
and under the basis of Eq.~\eqref{eq:PJT_basis} are expressed in matrix form as
\begin{equation}\label{eq:sigmasDJT}
\begin{aligned}
\bar{\sigma}_z &= \frac{1}{2}\begin{bmatrix}
-1 & 0 & 1 \\
0 & 2 & 0 \\
1 & 0 & -1
\end{bmatrix}\text{,} &
\bar{\sigma}_x &= \frac{1}{\sqrt{2}}\begin{bmatrix}
0 & -1 & 0 \\
-1 & 0 & 1 \\
0 & 1 & 0
\end{bmatrix}\text{.}
\end{aligned}
\end{equation}
Furthermore, based on basis of Eq.~\eqref{eq:PJT_basis} and 
taking Eq.~\eqref{eq:mix_C} into consideration, the effective DJT Hamiltonian is
\begin{equation}\label{eq:HDJTeffective}
\hat{H}_{\text{DJT}}^{\text{eff}} = (1-C^2) F_2(\bar{\sigma}_z\hat{X} - \bar{\sigma}_x\hat{Y})\text{.}
\end{equation}

The electron-phonon coupling $\widetilde{F}$ in PJT is about twice that of $F$ in DJT, which due to the double $e$ orbitals are JT unstable of $|xx\rangle$ in ($ee$) configuration. In contrast, in the ($ae$) configuration, only one $e$ orbital is JT unstable. Finally, combining the PJT and DJT and electron-electron interaction, the final effective electron-phonon coupling Hamiltonian $\hat{H}_{\mathrm{el-ph}}^{\mathrm{eff}}$ of the shelving singlet state is
\begin{align}\label{eq:Hamiltonianeff}
\hat{H}_{\mathrm{el-ph}}^{\mathrm{eff}} &= C^2 \cdot (2F_2) \cdot (\hat{\sigma}_z \hat{X} - \hat{\sigma}_x \hat{Y}) \nonumber \\
&\quad + (1 - C^2) \cdot F_2 \cdot (\bar{\sigma}_z \hat{X} - \bar{\sigma}_x \hat{Y})\text{,}
\end{align}
where $C^2$ represents the contribution that is affected by the PJT effect and induces ISC through the $\Gamma_z$ parameter. Similarly, the $(1-C^2)$ contribution is governed by DJT and induces ISC by means of $\Gamma_\pm$. The full Hamiltonian for the $\ket{{}^1\widetilde{E}} \oplus \ket{{}^1\widetilde{A}_1}$ system is
\begin{align}
\hat{H}&=\hat{H}_e+\hat{H}_{\mathrm{osc}}+\hat{H}_{\mathrm{el-ph}}^{\mathrm{eff}}\text{.} \label{eq:full_LBISC}
\end{align}
In this work, we mainly focus on the ISC transition from $\ket{{}^1\widetilde{E}}$ to $\ket{{}^3{A_2}}$. Based on Eq.~\eqref{eq:full_LBISC}, the $e$ phonon modes expansion could result in the following vibronic wavefunctions of
\begin{align}
|\widetilde{\Psi}\rangle&=\sum_{n,m}^\infty \Big[c_{nm}^{xx}|xx\rangle\otimes|nm\rangle+c_{nm}^{xy}|xy\rangle\otimes|nm\rangle\notag\\
&\quad+c_{nm}^{yy}|yy\rangle\otimes|nm\rangle\Big]\text{,} \label{eq:PsiTilde}
\end{align}
where the expansion of the phonon modes in the Born--Oppenheimer basis $\displaystyle{\ket{nm}=\frac{1}{\sqrt{nm}}(a_x^\dagger)^n(a_y^\dagger)^m\ket{00}}$ is limited to 10, i.e., $(n + m \le 10)$ to satisfy numerical convergence. Then, the expression of the $\ket{{}^1\widetilde{E}_\pm}$ is
\begin{align}\label{eq:E_pm_vibr}
\ket{{}^1\widetilde{E}_\pm} &= \sum_{i=1}^\infty \left( c'_i \ket{{}^1\bar{E}_\pm} \otimes \ket{\chi_i(A_1)} + d'_i \ket{{}^1A_1} \otimes \ket{\chi_i(E_\pm)} \right. \nonumber \\
&\quad \left. + f'_i \ket{{}^1\bar{E}_\mp} \otimes \ket{\chi_i(E_\mp)} + g'_i \ket{{}^1\bar{E}_\pm} \otimes \ket{\chi_i(A_2)} \right)\text{.}
\end{align}
Similarly to $|\varrho_i(...)\rangle$ in Eq.~\eqref{eq:ISC_upper}, the $|\chi_i(...)\rangle$ in Eq.~\eqref{eq:E_pm_vibr} also depicts symmetry-adapted vibrational wavefunctions. The ISC transition from $\ket{{}^1\widetilde{E}}$ to $\ket{{}^3{A_2}}$ is one kind of SOC-driven scattering, which is mediated by the electron-phonon interactions. In these two centers, also in the NV-diamond, the energy gap between $\ket{{}^1\widetilde{E}}$ and $\ket{{}^3{A_2}}$ are far larger than the strength of SOC, indicating that the electrons will be scattered to the vibration levels $\bra{\ldots}$ of $\ket{{}^3{A_2}}$ ground state. During this process, the $E$ phonons play a vital role arising from the PJT and DJT effects. In the upper branch of ISC discussed in Section~\ref{sec:ISC_upper}, we assume that the SOC would not change significantly during the transition process. Hence, the SOC data is also consistent with the upper branch in Section~\ref{sec:ISC_upper}. The ISC rate could also be expressed by the variety of Fermi's golden rule~\cite{thiering2018theory} like Eq.~\eqref{eq:Gamma_1A1_upper}. 
However, we note that the ISC transitions mechanism towards $\ket{^{3}A_{2}^{0}}$ and $\ket{^{3}A_{2}^{\pm}}$ are different. 

The $\Gamma_z$ between $\ket{{}^1\widetilde{E}}$ and $\ket{^{3}A_{2}^{0}}$ could be expressed as
\begin{align}\label{eq:LB_Gamma_z}
\Gamma_z &= \frac{2\pi C^2}{\hbar} \sum_{\ket{\ldots}} \vert \bra{\ldots} \otimes \bra{{}^3A_2^0} \hat{W} \ket{{}^1\widetilde{E}} \vert ^2 \delta(\Sigma - E(\ket{\ldots})) \nonumber \\
&= \frac{2\pi C^2}{\hbar} \sum_i^\infty 4\lambda_z^2 d'_i{}^2 \left| \braket{\ldots | \chi_i(E_\pm)} \right|^2 \delta(\Sigma - n_i \hbar \omega_E) \nonumber \\
&\approx \frac{8\pi \lambda_z^2 C^2}{\hbar} \sum_i^\infty d_i'^2 S_E^{(n_i)}(\Sigma) \nonumber \\
&= \frac{8\pi \lambda_z^2 C^2}{\hbar} F_E(\Sigma)\text{,}
\end{align}
where the $d'_i$ coefficient means the contribution of $\ket{^{1}A_{1}}$ in $\ket{{}^1\widetilde{E}}$, which connects to $\ket{^{3}A_{2}^{0}}$ by the $\lambda_z$~\cite{maze2011properties,doherty2011negatively,thiering2018theory}; $n_i$ represents the $i$-th $\ket{\chi_i(E_\pm)}$ vibronic function. $F_E$ is the PJT-modulated phonon overlap function based on the phonon overlap spectral function $S_E$. $\Sigma$ is the energy gap between $\ket{{}^1\widetilde{E}}$ and $\ket{^{3}A_{2}}$ as shown in Fig.~\ref{fig:fifteen_levels}. A recursive formula was used to avoid discrete quantum energy levels, causing the overlap in $F_E$ to be zero~\cite{thiering2018theory}. Except for the $\Gamma_z$, there are also $\Gamma_{\pm}$ and $\Gamma_{\mp}$ between $\ket{{}^1\widetilde{E}}$ and $\ket{^{3}A_{2}^{\pm}}$ driven by $\lambda_{\perp}$ with a form as
{\small
\begin{align}\label{eq:LB_Gamma_pm}
\Gamma_{\pm} &= \frac{2\pi (1-C^2)}{\hbar} \sum_{\ket{\ldots}} \vert \bra{\ldots} \otimes \bra{{}^3A_2^{\pm}} \hat{W} \ket{{}^1\widetilde{E}} \vert^2 \delta(\Sigma - E(\ket{\ldots})) \nonumber \\
&= \frac{2\pi (1-C^2)}{\hbar} \sum_i^\infty \lambda_{\bot}^2 c'_i{}^2 \left| \braket{\ldots|\chi_i(A_1)} \right|^2 \delta(\Sigma - n_i \hbar \omega_E) \nonumber \\
&\approx \frac{2\pi \lambda_{\bot}^2 (1-C^2)}{\hbar} \sum_i^\infty c_i'^2 S_E^{(n_i)}(\Sigma) \nonumber \\
&= \frac{2\pi \lambda_{\bot}^2 (1-C^2)}{\hbar} F'_E(\Sigma)\text{,}
\end{align}
}
and
{\small
\begin{align}\label{eq:LB_Gamma_mp}
\Gamma_{\mp} &= \frac{2\pi (1-C^2)}{\hbar} \sum_{\ket{\ldots}} \vert \bra{\ldots} \otimes \bra{{}^3A_2^{\pm}} \hat{W} \ket{{}^1\widetilde{E}} \vert^2 \delta(\Sigma - E(\ket{\ldots})) \nonumber \\
&= \frac{2\pi (1-C^2)}{\hbar} \sum_{i=1}^{\infty} \lambda_{\bot}^2 f'_i{}^2 \left| \braket{\ldots | \chi_i(A_1)} \right|^2 \delta(\Sigma - n_i \hbar \omega_E) \nonumber \\
&\approx \frac{2\pi \lambda_{\bot}^2 (1-C^2)}{\hbar} \sum_{i=1}^{\infty} f'_i{}^2 S_E^{(n_i)}(\Sigma) \nonumber \\
&= \frac{2\pi \lambda_{\bot}^2 (1-C^2)}{\hbar} F''_E(\Sigma)\text{,}
\end{align}
}
where $F'_E$ and $F''_E$ represent the phonon overlap spectral functions resulting from the DJT effect. 

The $C^2$ parameter could be obtained numerically by the character of the KS wavefunctions of the calculated closed-shell $|xx\rangle$, which results in the contribution of the $a$ KS orbital in the two-particle wave functions~\cite{thiering2018theory}. When labeling the true (mixed) KS state as $|\xi\xi\rangle$ and the contribution of $a$ and $e_x$ by $s$ and $p$, then we get
\begin{align}
|\xi\xi\rangle &= (p|e_x\rangle + s|a\rangle)(p|e_x\rangle + s|a\rangle) \notag \\
&= p^2|e_x e_x\rangle + \sqrt{2}ps \left( \frac{|ae_x\rangle + |e_xa\rangle}{\sqrt{2}} \right)_{|^1E'_x\rangle} \notag \\
&\quad + s^2 \left( |aa\rangle \right)_{|^1A'_1\rangle}\text{,} \label{eq:xx_components}
\end{align}
where $p$ and $s$ means the respective contribution of $e_x$ and $a$; and $(C^2=1-2p^2s^2)$ could be read out directly. The $F_2$ in Eq.~\eqref{eq:Hamiltonianeff} could be obtained by the relationship of
\begin{equation}\label{eq:F2}
    F_2 = \frac{\sqrt{2\hbar\omega_{E}E_{\text{JT2}}}}{1 + C^2}\text{,}
\end{equation}
where the effective phonon mode $\hbar\omega_{E}$ and JT energy $E_{\text{JT2}}$ arise from the fitting and energy of the distorted $|xx\rangle$ geometry, respectively~\cite{thiering2018theory}. All the obtained parameters of Eq.~\eqref{eq:Hamiltonianeff} are shown in Table~\ref{tab:PJT}. Finally, the calculated $c'_i{}^2$, $d'_i{}^2$ and $f'_i{}^2$ coefficients are shown in Table~\ref{tab:all_coeff} of Appendix~\ref{appendix:p_q}. 

\begin{figure}[h] 
\includegraphics[width=\linewidth]{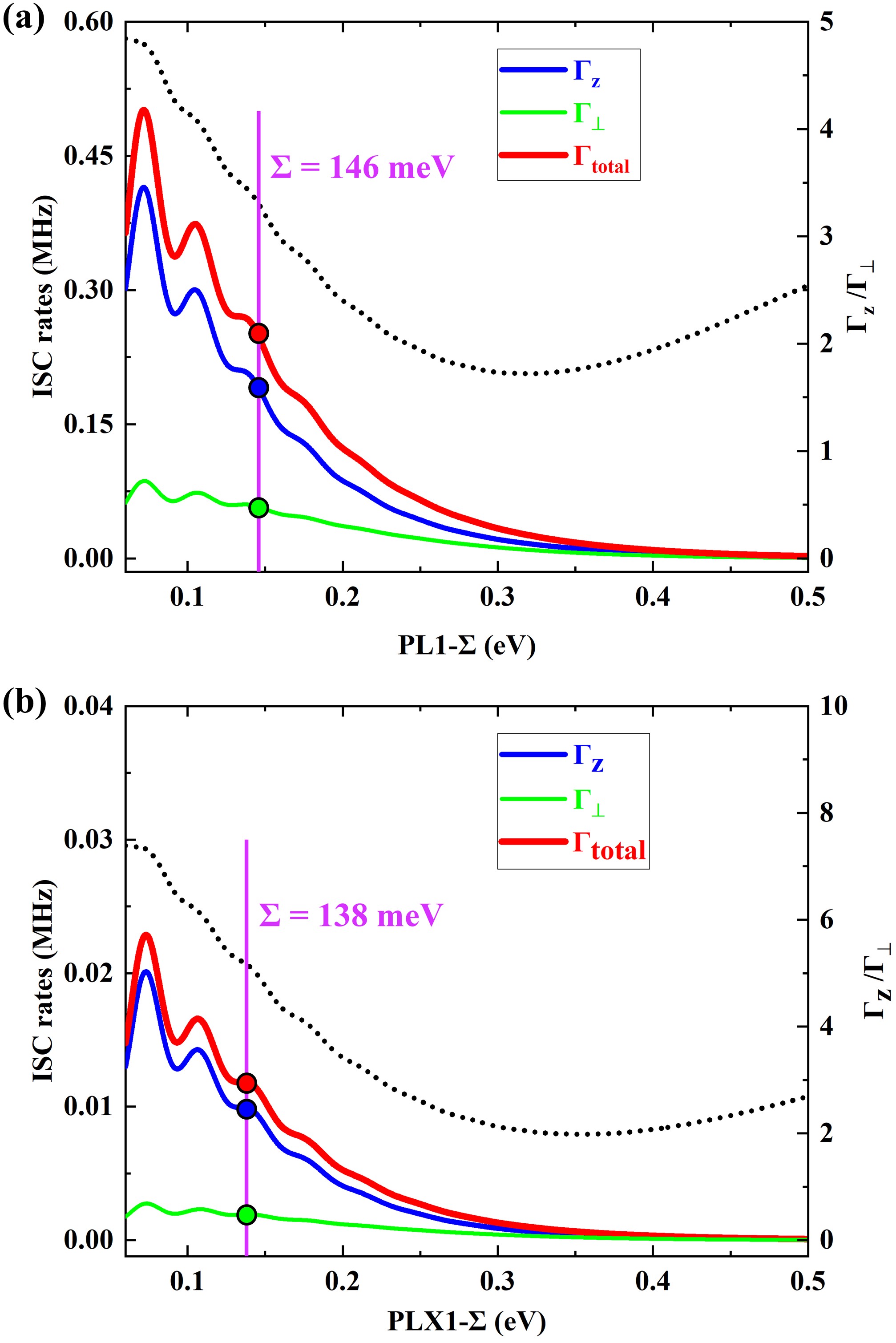}
\caption{\label{fig:lower_all} The calculated rates of lower branch and $(\Gamma _z/\Gamma _ \bot )$ ratios. $\Gamma_\bot = \Gamma_\pm + \Gamma_\mp$ and $\Gamma_\text{total} = \Gamma_\text{z} + \Gamma_\bot$. $\Sigma$ is the splitting between triplet $\ket{^{3}A_{2}}$ and degenerate singlet $\ket{{}^1E_{\mp}}$ as shown in Fig.~\ref{fig:fifteen_levels}. The details of the $\Sigma$ values are shown in Table~\ref{tab:PJT}.}
\end{figure}

The lower branch of ISC transition occurs from the double degenerate $\ket{^1E}$ state to the $\ket{{}^3A_2}$ triplet state, in contrast to the upper branch, which transitions from triplet to singlet states. Both the axial and transverse components of SOC are involved in this process: ${\Gamma _z}$ (from $\ket{^1E}$ to $\ket{^{3}A_{2}^{0}}$) is associated with $\lambda_z$ and ${\Gamma _ \bot}$ (from $\ket{^1E}$ to $\ket{^{3}A_{2}^{\pm}}$) is associated with $\lambda_ \bot$. The ${\Gamma _z / \Gamma _ \bot }$ ratio is also crucial and could be advantageous to quickly assessing the spin polarizability. Additionally, the ratio highly depends on the combined nature of DJT and PJT effects with SOC. The calculated lower branch rates are illustrated in Fig.~\ref{fig:lower_all}. For the PL1 center, ${\Gamma _z = 0.19}$~MHz and ${\Gamma _ \bot = 0.06}$~MHz at $\Sigma_{\rm{PL1}} = 146$~meV, where the related ratio of ${\Gamma _z}/{\Gamma _ \bot }= 3.30$. For the PLX1 center, as shown in Fig.~\ref{fig:lower_all}(b), ${\Gamma _z = 0.01}$~MHz and ${\Gamma _ \bot = 0.002}$~MHz at $\Sigma_{\rm{PLX1}} = 138$~meV, where the related ratio of ${\Gamma _z}/{\Gamma _ \bot }= 5.17$.  By comparing the differences between the theoretical and experimental SOC results of the PL1 center, we conclude that the calculated SOC of the PLX1 center may be slightly underestimated, resulting in a smaller rate than the actual value.
\begin{table}[h]
\caption{\label{tab:PJT}PJT effect parameters of PL1 and PLX1 centers. All data are in meV except for $C^2$.}
\begin{ruledtabular}
\renewcommand{\arraystretch}{1.2}  
\begin{tabular}{lccccc}  %
      & $\Lambda_e$ & $\hbar\omega_{E}$ & $E_{\mathrm{JT2}}$ & $C^2$ & ${F_2}$  \\
\hline 
PL1   & 847 & 36.6 & 118.4 & 0.89 & 49.3 \\
PLX1  & 891 & 39.5 & 109.5 & 0.91 & 48.7 \\
\end{tabular}
\end{ruledtabular}
\end{table}

\subsection{\label{sec:summary_outcomes}PL lifetime and ODMR contrast}
\begin{figure*}    
\includegraphics[width=\linewidth]{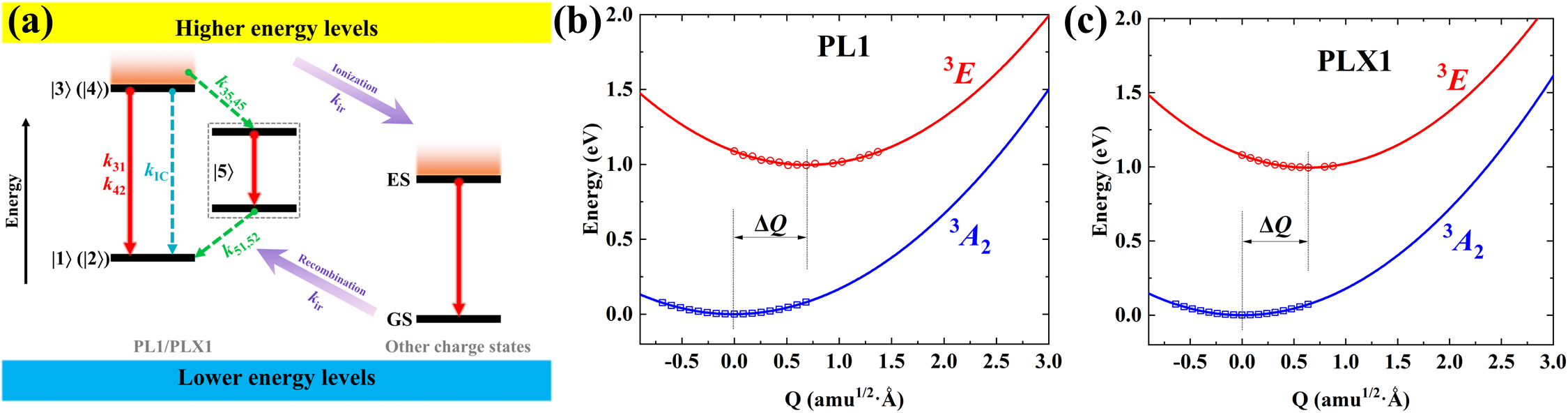} 
\caption{\label{fig:five_level_k_IC}Diagram of the five-level rate-equation model with major associated rates. Key features: black line (energy levels), glowing arrow (radiative transitions), dashed line (phonon-mediated nonradiative transitions), and gradient arrow line (ionization/recombination transitions). GS (ES) is the ground (excited) state. $\Ket{1} = \Ket{{^3A_2^0}}$, $\Ket{2} = \Ket{{^3A_2^\pm}}$, $\Ket{3 (4)}$ is sublevel of  $\ket{{}^3E}$ with $m_s = 0 (\pm 1)$. Gradients above the energy levels indicate phonon sidebands. Charge states vary while maintaining the same atomic structure. Transition rates between states $\Ket{i}$ and $\Ket{j}$ are denoted as $k_{ij}$. Panels (b) and (c) illustrate energy level crossings between $\ket{{}^3A_2}$ and $\ket{{}^3E}$ for the PL1 and PLX1 centers at 0 K, showing energy $E$ versus configuration coordinate $Q$. The vertical offset of potential energy surface $\Delta Q$ shows the values of $0.703 (0.638) \sqrt{\text{amu}}\!\cdot\!\text{\AA}$ for PL1 (PLX1) center.}
\end{figure*}

The PL lifetime $\tau_{\text{PL}}$ of the $\ket{{}^3E}$ excited state is the reciprocal of the transition rate $k_{\text{PL}}$ from the $\ket{{}^3E}$ to the $\ket{{}^3A_2}$ ground state. There are various pathways form $\ket{{}^3E}$ to $\ket{{}^3A_2}$, mainly consisting of the radiative transition $k_{\text{rad}}$, the nonradiative transition $k_{\text{ph}}$, and the ionization (recombination) transition $k_{\text{ir}}$, with the relationship as
\begin{equation}
\frac{1}{\tau_{\text{PL}}} = k_{\text{PL}} = k_{\text{rad}} + k_{\text{ph}} + k_{\text{ir}}.
\label{eq:PL_lifetime}
\end{equation}
Based on the above results, for the sake of simplicity, energy levels shown in Fig.~\ref{fig:fifteen_levels} are enumerated into a five-level rate-equation model with major transition rates as shown in Fig.~\ref{fig:five_level_k_IC}(a). The radiative transition is a spin-conserving transition with photon emission dominated by the selection rules, which is mostly the source of fluorescence signals in QIS experiments and named as $k_{31}$ and $k_{42}$ in Fig.~\ref{fig:five_level_k_IC}(a). The thermally assisted nonradiative transition contains two parts. One is the spin-conserving direct decay from $\ket{{}^3E}$ to $\ket{{}^3A_2}$, also called the internal conversion (IC) with a rate $k_{\text{IC}}$~\cite{nizovtsev2001modeling}. The other one is the ISC transition with a rate of $k_{\text{ISC}}$, which is the composite transition of upper ($k_{35}$ and $k_{45}$ in Fig.~\ref{fig:five_level_k_IC}(a)) and lower ($k_{51}$ and $k_{52}$ in Fig.~\ref{fig:five_level_k_IC}(a)) branches. Ignoring any weaker phonon-mediated transitions, $k_{\text{ph}} = k_{\text{IC}} + k_{\text{ISC}}$ in general. Under some specific conditions, the ionization (recombination) transition via other charge states~\cite{yuan2020charge} is also non-negligible, which also affects the ODMR contrast. 

The radiative transition is the spin-conserved dipole transition between the $\ket{{}^3E}$ excited state to the $\ket{{}^3A_2}$ ground state dominated by the selection rules. The expression of the transition rate $k_{\text{rad}}$ is
\begin{equation}
k_{\text{rad}} = \frac{n E_{\text{ZPL}}^3 \left| \mu \right|^2}{3\pi \epsilon_0 \hbar^4 c^3},
\label{eq:rad_decay_rate}
\end{equation}
where ${\varepsilon_0}$ is the vacuum permittivity; $\hbar$ is the reduced Planck constant; $c$ is the speed of light in vacuum; $n = 2.647$ is the refractive index of 4H-SiC~\cite{davidsson2020theoretical}; ${E_{\rm{ZPL}}}$ is the ZPL energy, and $\mu $ is the optical-transition dipole moment. Referring to Fig.~\ref{fig:five_level_k_IC}(a), $k_{\text{rad}}$ corresponds to two transitions: $k_{31}$ between the $\ket{0}$ sublevels of $\ket{{}^3E}$ and $\ket{{}^3A_2}$, and $k_{42}$ between the $\ket{\pm 1}$ sublevels of $\ket{{}^3E}$ and $\ket{{}^3A_2}$. We used the $\Delta$SCF method to calculate the ZPL values of 1.15~eV and 1.09~eV for the PL1 and PLX1 centers. These values are in line with experimental results of 1.095~eV (1132~nm) for the PL1 center~\cite{falk2013polytype}, and 0.999~eV (1241~nm) for the PLX1 center~\cite{zargaleh2016evidence}. Furthermore, the dipole moments were calculated using the pseudo wavefunctions of the $a$ and $e$ KS levels in JT-distorted excited state~\cite{mohseni2023positively}, yielding values of 7.36~D and 8.44~D for PL1 and PLX1 centers. Finally, the calculated $k_{\text{rad}}$ values are 35.6 and 40.3~MHz for PL1 and PLX1 centers, corresponding to radiative times of 28.1 and 24.8~ns, respectively.

For the PL1 center, the radiative lifetime of $k_{\text{rad}} = 28.09$~ns aligns closely with another simulated value of 23.01~ns~\cite{davidsson2020theoretical}, as well as with the experimental value of around 16~ns at low temperature for both single and ensemble~\cite{crook2020purcell,falk2014electrically}. We also calculated the levels crossing between $\ket{{}^3A_2}$ and $\ket{{}^3E}$ as shown in Fig.~\ref{fig:five_level_k_IC}(b). The offset vertical $\Delta Q$ shows a value of $0.703 \sqrt{\text{amu}}\!\cdot\!\text{\AA}$ for PL1 center. From Fig.~\ref{fig:five_level_k_IC}(b), even in a large configuration coordinate ($Q$) of 3 $\sqrt{\text{amu}}\!\cdot\!\text{\AA}$, there is still no crossing point between the two levels. It is because of the large energy gap of about 1.1~eV between $\ket{{}^3A_2}$ and $\ket{{}^3E}$~\cite{li2022room}, which results in almost no overlap between their APESs. Hence, the $k_{\text{IC}}$ for the PL1 center could be ignored compared to the $k_{\text{rad}}$ at 0~K or low temperature. 

However, the calculated radiative lifetime of 24.81~ns for the PLX1 center deviates from the experimental measured excited state lifetime of 2.7~ns at low temperature~\cite{wang2020coherent}. From Fig.~\ref{fig:five_level_k_IC}(c), there is also no overlap between APESs of $\ket{{}^3E}$ to the $\ket{{}^3A_2}$ because of a large energy gap of 1~eV~\cite{zargaleh2016evidence}, which means the $k_{\text{IC}}$ at 0~K of PLX1 center also could be ignored like for PL1 center. 

Therefore, this deviation should be mainly caused by the non-negligible $k_{\text{ir}}$. In the case of NV-diamond, which is isovalent center to the PLX1 center, $k_{\text{ir}}$ has a significant maximum ionization rate of 21.2~MHz and a recombination rate of 390.3~MHz under certain circumstances~\cite{yuan2020charge}. Besides, the ionization/recombination transition depends heavily on complicated conditions such as laser power and sequence, surface or internal impurities, and readout protocol~\cite{barry2020sensitivity}. Ref.~\onlinecite{anderson2022five} also demonstrates a single-shot readout of the PL1 center via spin-to-charge conversion. The competing non-negligible $k_{\text{ir}}$ in special circumstances shows excellent reference significance for the PLX1 center, which reasonably explains the deviation between the theoretical and experimental results.
\begin{table}
\caption{\label{tab:all_outcome} Summary of the calculated major rates in Fig.~\ref{fig:five_level_k_IC} for PL1 and PLX1 centers. The $k_{ij}$ means rates shown in Fig.~\ref{fig:five_level_k_IC}, and $\Gamma$ is shown in Fig.~\ref{fig:fifteen_levels}. $k_{45} =  \left( \Gamma_{A_1} + 2\Gamma_{E_{1,2}} + \Gamma_{A_2} \right)/4$ is for the case of off-resonant excitation.}
\begin{ruledtabular}
\renewcommand{\arraystretch}{1.2}
\begin{tabular}{ccc}
Parameters                  & PL1 (MHz) & PLX1 (MHz) \\
\hline 
$k_{31}$ ($k_{42}$)         & 35.60     & 40.31 \\
$\Gamma_{A_1}$              & 13.60     & 0.95 \\
$\Gamma_{E_{1,2}}$          & 6.85      & 0.03 \\
$\Gamma_{A_2}$              & 0.46      & $\approx 0$ \\
$k_{45}$                    & 6.94      & 0.25 \\
$k_{51}$ ($\Gamma_z$)       & 0.19      & 0.01 \\
$k_{52}$ (${\Gamma _\bot}$) & 0.06      & $\approx 0$ \\
\end{tabular}
\end{ruledtabular}
\end{table}

Finally, all the resulting major transition rates related to ODMR contrast are collected and shown in Table~\ref{tab:all_outcome}. Besides, the average arises from the assumption that one event only occurs from one of the vibronic states and occupation of any of the triplets can happen with the same probability when the electron is excited from the $\ket{m_s = \pm 1}$ sub-state of the ground state at low temperatures. The quantum yield is a key indicator for evaluating quantum information readout efficiency~\cite{ping2021computational}, with a relation of $\eta_\text{QY} = k_\text{rad}/k_\text{PL}$. Based on Eq.~\eqref{eq:PL_lifetime} and data shown in Table~\ref{tab:all_outcome}, the resulting $\eta_\text{QY}$ are $83.86 {\kern 1pt} \%$ and $99.41 {\kern 1pt} \%$ for PL1 and PLX1 centers, respectively. Though the actual pulse off-resonant ODMR readout contrast $\mathcal{C}$ depends on many factors, the trend can be simplified to an expression with defect intrinsic parameters and expressed as~\cite{li2022room, gali2019ab}
\begin{equation} \label{eq:contrast}
\mathcal{C} = \frac{\tau_{\pm 1} - \tau_0}{\tau_0} = \frac{k_0 - k_{\pm 1}}{k_{\pm 1}}\text{,}
\end{equation}
where the $\tau_{\pm 1}$ and $\tau_0$ are optical lifetimes of the $\ket{\pm 1}$ and $\ket{0}$ for the excited state $\ket{{}^3E}$ and inverse of the rates $k_0$ and $k_{\pm 1}$, respectively. Based on Fig.~\ref{fig:five_level_k_IC}, there should have $k_0 = k_{31} + k_{\text{IC}} + k_{35} + k_{\text{ir}}$, and $k_{\pm 1} = k_{42} + k_{\text{IC}} + k_{45} + k_{\text{ir}}$. $k_{\text{rad}} = k_{31} = k_{42}$~\cite{li2022room}. Besides, the nonradiative transition rate of $k_{\text{35}}$ is extremely weak, so that can be ignored~\cite{gali2019ab,li2022room}. Based on the results shown in Fig.~\ref{fig:five_level_k_IC}(b) and (c), the direct nonradiative transition rate $k_{\text{IC}}$ could also be neglected. Therefore, $k_0 = k_{31} + k_{\text{ir}}$ and $k_{\pm 1} = k_{42} + k_{45} + k_{\text{ir}}$. For the PL1 center, based on the comparison and analysis of the above calculated and experimentally measured radiative lifetimes, we concluded that under normal circumstances, the contribution of $k_{\text{ir}}$ is very small and can also be ignored there. Taking data shown in Table~\ref{tab:all_outcome}, the ideal ODMR contrast $\mathcal{C} = -16.31 {\kern 1pt} \%$.  For the PLX1 center without considering $k_{\text{ir}}$, based on parameters shown in Table~\ref{tab:all_outcome}, the resulted contrast is $-0.6 {\kern 1pt} \%$. As mentioned above, the ionization (recombination) transition (with rate of $k_{\text{ir}}$) may play an important role in the whole transition loop, which results in the deviation between the calculated radiative lifetime and the experimental measured excited state lifetime when ignoring the $k_\text{IC}$ and precipitously decrease the contrast to $-0.11 {\kern 1pt} \%$~\cite{wang2020coherent}. 

The ideal contrast calculated by Eq.~\eqref{eq:contrast} rests upon several basic assumptions~\cite{li2022room}. However, real-world experiments are imperfect and can result in a lower contrast (in absolute values) than predicted. This deviation from the theoretical limit is influenced by factors such as setup performance, sample preparation, pulse sequence design, and the test environment. Eq.~\eqref{eq:contrast} provides an upper bound for the pulsed microwave ODMR contrast that can be principally approached by optimizing the samples, optical and microwave controls, and protocols in the experiments. Additionally, this theoretical framework is broadly transferable to other defect centers with accessible optical spin polarization loops. The ODMR contrast arises from differing transition rates of different spin states from ES to GS and primarily depends on non-radiative ISC transitions. For instance, Fig.~\ref{fig:five_level_k_IC}(a) clearly displays the major transition pathways to jointly determine the ODMR contrast of a $C_\text{3v}$ system with a total spin of 1, and the model can be easily transplanted to any system with the same symmetry and spin state. As for another total spin system, such as 3/2, the essence of ODMR remains unchanged, and this work still has strong reference significance.

\section{\label{sec:conclusion}Conclusion}
In this work, we comprehensively investigate the microscopic magneto-optical properties and optical spin polarization of PL1 and PLX1 centers in 4H-SiC for potential qubit applications in quantum information science by employing the first-principles calculations. First, we present a detailed overview of the KS levels with a quantitative sketch, followed by a thorough analysis of the two-particle basis functions within the hole notation and their intrinsic hierarchy. The DJT-reduced SOC is fully investigated to deduce the ISC transition rates. Additionally, the ZFS among different spin sublevels of both the ground and excited triplet states is computed, which provides key parameters of the ODMR protocol and, along with the parallel component of SOC, determines the fine energy structure. Moreover, the upper and lower branches of the ISC transition mediated by the electron-phonon coupling are well demonstrated, particularly unveiling the JT nature in different cases. The PL lifetime of excited triplet states is also investigated in detail. Finally, based on the calculated major transition rates among key states, an optical spin polarization loop is fully assembled, and the optimal ODMR contrast is derived in detail. This work reveals the electron-phonon coupling mechanism underlying the optical spin polarization. Our results imply that there is potential to optimize the ODMR contrast of the PL1 and PLX1 centers, which is of high importance to versatile applications such as increasing the sensitivity in quantum sensing.

\begin{acknowledgments}
Support by the Ministry of Culture and Innovation and the National Research, Development and Innovation Office within the Quantum Information National Laboratory of Hungary (Grant No.\ 2022-2.1.1-NL-2022-00004) is much appreciated. AG acknowledges the high-performance computational resources provided by KIF\"U (Governmental Agency for IT Development) institute of Hungary and the European Commission for the projects QuMicro (Grant No.\ 101046911) and SPINUS (Grant No.\ 101135699).

\end{acknowledgments}

\appendix

\section{\label{appendix:SOC}SOC calculation details}

\begin{table*} 
\caption{Coefficients utilized in both upper and lower branches.}
\label{tab:all_coeff}
\begin{ruledtabular}
\renewcommand{\arraystretch}{1.1}
\begin{tabular}{ccccccccccccc}
\multirow{2}{*}{$i = m + n$} & \multicolumn{6}{c}{PL1} & \multicolumn{6}{c}{PLX1} \\
\cmidrule(lr){2-7} \cmidrule(lr){8-13}
& $c_i^2$ & $c'_i{}^2$ & $d_i^2$ & $d'_i{}^2$ & $f_i^2$ & $f'_i{}^2$ & $c_i^2$ & $c'_i{}^2$ & $d_i^2$ & $d'_i{}^2$ & $f_i^2$ & $f'_i{}^2$ \\  
\hline
0 & 0.274 & 0.000 & 0.000 & 0.000 & 0.000 & 0.940 & 0.301 & 0.960 & 0.000 & 0.000 & 0.000 & 0.000 \\
1 & 0.000 & 0.027 & 0.328 & 0.007 & 0.000 & 0.000 & 0.000 & 0.000 & 0.334 & 0.006 & 0.000 & 0.017 \\
2 & 0.201 & 0.013 & 0.013 & 0.000 & 0.000 & 0.010 & 0.186 & 0.007 & 0.016 & 0.000 & 0.000 & 0.009 \\
3 & 0.010 & 0.000 & 0.093 & 0.000 & 0.010 & 0.001 & 0.011 & 0.000 & 0.080 & 0.000 & 0.011 & 0.000 \\
4 & 0.029 & 0.000 & 0.013 & 0.000 & 0.000 & 0.000 & 0.023 & 0.000 & 0.014 & 0.000 & 0.000 & 0.000 \\
5 & 0.003 & 0.000 & 0.010 & 0.000 & 0.003 & 0.000 & 0.003 & 0.000 & 0.008 & 0.000 & 0.003 & 0.000 \\
$\cdots$ & $\cdots$ & $\cdots$ & $\cdots$ & $\cdots$ & $\cdots$ & $\cdots$ & $\cdots$ & $\cdots$ & $\cdots$ & $\cdots$ & $\cdots$ & $\cdots$ \\
Sum & 0.519 & 0.041 & 0.460 & 0.008 & 0.012 & 0.951 & 0.526 & 0.967 & 0.454 & 0.007 & 0.013 & 0.026 \\
\end{tabular}
\end{ruledtabular}
\end{table*}

Based on our previous work~\cite{thiering2017ab}, we find that a supercell size of \(6 \times 6 \times 2\) with 576 atoms may not adequately ensure the convergence of the SOC value. Therefore, variations of supercell size are necessary for finding the trend and achieving convergence of ${\lambda _z}$. Unlike cubic diamond crystals, 4H-SiC belongs to a hexagonal crystal system, making it unsuitable for extending the supercell equally in the $x$, $y$, and $z$ directions as is executed for NV-diamond~\cite{thiering2017ab}. As shown in Table~\ref{tab:all_soc}, we expand lattice constants $a$, $b$ together as ($5 \times 5$), ($6 \times 6$), ($7 \times 7$), and ($8 \times 8$), while expanding $c$ independently as 1, 2, 3, and 4 (representing integer multiples of the unit cell lattice constant). 

For all supercell sizes, we perform SOC calculations relying on the PBE functional~\cite{perdew1996generalized,steiner2016calculation}, which makes it possible to calculate ultra-large supercells such as $8 \times 8 \times 4$ with 2046 atoms. Besides, all supercell sizes shown in Table~\ref{tab:all_soc} also ensure sufficient accuracy for calculating ${\lambda_z}$ based on the PBE functional. The strength of ${\lambda_z}$ arises from the splitting of the two double-degenerate $e$ levels when both levels are half occupied. The total energy is converged to $10^{-7}$~eV with a fixed high-symmetry geometry. All calculated SOC results are summarized in Table~\ref{tab:all_soc}.

\begin{figure} [t]
\centering
\includegraphics[width=\linewidth]{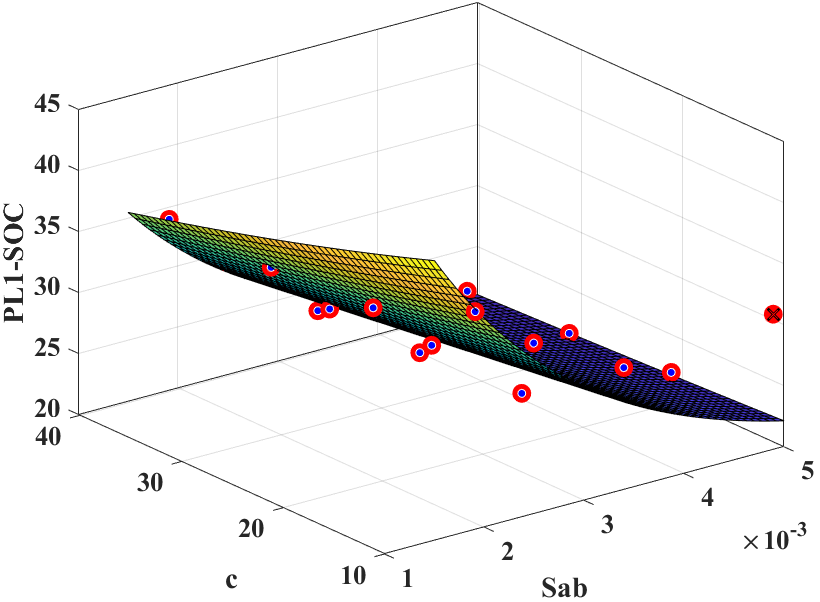}
\caption{SOC fitting of PL1 center based on the data in Table~\ref{tab:all_soc} and Eq.~\eqref{eq:fit_lambda_z}.}
\label{fig:SOC_fitting_PL1}
\end{figure}

From Table~\ref{tab:all_soc}, we find that the value of ${\lambda _z}$ changes differently when the supercell size changes along the $c$-axis compared to when it changes along the $a$ and $b$ axes simultaneously and synchronously. Therefore, it is necessary to fit the various results using a composite function that can contain two trends at the same time and obtain a convergence value to reflect the characteristics of an isolated qubit. Unlike in NV-diamond, the extension of the color center's wave function along the transverse ($a$ and $b$ axes) and axial ($c$ axis) directions is not identical in 4H-SiC. Hence, the most practical fitting function is composite exponential functions with an expression as
\begin{equation}
\lambda_{z}\left(x_{ab},y_c\right)=A\cdot \exp\left(B\cdot x_{ab}+C\cdot y_c\right)+\lambda_{z0},
\label{eq:fit_lambda_z}
\end{equation}
where $A$, $B$, $C$, and $\lambda_{z0}$ are fitting parameters; $x_{ab} = \displaystyle{\frac{2}{\sqrt{3} \, ab}}$ represents the reciprocal of the supercell cross-sectional area, $y_c = c$, and $\lambda_{z0}$ corresponds to the SOC for an isolated qubit. The fitting visualization of the PL1 center is displayed in Fig.~\ref{fig:SOC_fitting_PL1}, where the point with $\times$ is a bad point that is removed for better fitting. The fitting of the PLX1 center is similar to the PL1 center. The final fitted convergent $\lambda_{z0}$ is 18.501 and 9.664~GHz for PL1 and PLX1 centers, respectively.

\begin{table} [h]
\caption{\label{tab:all_soc} Calculated $\lambda _z$ values of all cases of supercell size.}
\begin{ruledtabular}
\renewcommand{\arraystretch}{1.2} 
\begin{tabular}{ccccc}
\begin{tabular}[c]{@{}c@{}}Supercell size\\ \hline $a \times b \times c$ \end{tabular} 	&	Volume ($\text{\AA}^3$) & No. of atoms	& \begin{tabular}[c]{@{}c@{}}${\lambda_z}$ (GHz)\\ \hline PL1 / PLX1 \end{tabular}   \\
\hline
$5 \times 5 \times 1$ & 2087.98 & 198 & 31.07 / 22.00 \\
$5 \times 5 \times 2$ & 4175.96 & 400 & 22.49 / 17.65 \\
$5 \times 5 \times 3$ & 6263.94 & 598 & 21.88 / 17.77 \\
$5 \times 5 \times 4$ & 8351.92 & 798 & 21.52 / 17.65 \\
$6 \times 6 \times 1$ & 3006.69 & 288 & 29.98 / 25.03 \\
$6 \times 6 \times 2$ & 6013.39 & 576 & 24.06 / 23.45 \\
$6 \times 6 \times 3$ & 9020.08 & 864 & 23.58 / 23.33 \\
$6 \times 6 \times 4$ & 12026.77 & 1152 & 23.21 / 23.33 \\
$7 \times 7 \times 1$ & 4092.44 & 390 & 33.97 / 31.31 \\
$7 \times 7 \times 2$ & 8184.89 & 784 & 29.98 / 31.55 \\
$7 \times 7 \times 3$ & 12277.33 & 1174 & 29.14 / 31.31 \\
$7 \times 7 \times 4$ & 16369.77 & 1566 & 29.14 / 31.19 \\
$8 \times 8 \times 1$ & 5345.23 & 510 & 37.84 / 35.79 \\
$8 \times 8 \times 2$ & 10690.46 & 1024 & 34.34 / 36.51 \\
$8 \times 8 \times 3$ & 16035.70 & 1534 & 33.85 / 36.51 \\
$8 \times 8 \times 4$ & 21380.93 & 2046 & 33.97 / 36.51 \\
\end{tabular}
\end{ruledtabular}
\end{table}

\section{\label{appendix:p_q}Electron-phonon coupling coefficients and the reduction factors}

The $p$ factor derived from the electron-phonon coupling coefficients implies a mixture between $\ket{E^+}$ and $\ket{E^-}$ of $\ket{^3E}$, which quenches the effective angular momentum~\cite{thiering2017ab}. The $p$ factor is obtained in the following initial form of $p=\sum_{nm}(c_{nm}^2-d_{nm}^2)$ in Eq.~\eqref{eq:DJT_expension}. Based on the rewritten symmetry adapted wavefunctions of Eq.~\eqref{eq:ISC_upper}, the reduction factors $p$ and $q$ are obtained from
\begin{equation} \label{eq:pq_expressions}
\begin{aligned}
p &= \sum_i \left(c_i^2 - d_i^2 + f_i^2\right), \quad 
q &= \sum_i \left(c_i^2 - f_i^2\right)
\end{aligned}
\end{equation}
where $c_i^2$, $d_i^2$, and $f_i^2$ are expansion coefficients by solving the electron-phonon Hamiltonian Eq.~\eqref{eq:DJT_Hmiltonian}. The expansion is limited up to six oscillator quanta ($m + n \le 6$) for numerical convergence of $< 1\%$, which is more than the NV-diamond ($m + n \le 4$)~\cite{thiering2017ab} because of the larger $E_{\rm{JT}}$ and smaller $\hbar\omega_e$ of the two centers than those of NV-diamond. Table~\ref{tab:all_coeff} displays the calculated values of $c_i$, $d_i$, and $f_i$. Still, in Table~\ref{tab:all_coeff}, we present the row only up to 5 since all coefficients are so small to be ignored when $i > 5$. The expressions of symmetry-adapted vibrational wavefunctions $|\varrho_i(\Gamma_i)\rangle$ in Eq.~\eqref{eq:ISC_upper} are shown in Table~\ref{tab:upper_phonon}. From Table~\ref{tab:all_coeff} we find that $|\varrho_i(E_{1,2})\rangle$ phonon functions has a minimum quantum number of $i = 1$ and $|\varrho_i(A_2)\rangle$ phonon function has a minimum quantum number of $i = 3$. Besides, the $p$ factor resulted from quadratic DJT with the barrier energy because it will be ${\rm{10 {\kern 1pt} \% }}$ smaller when using only the linear DJT approximation. Based on data displayed in Table~\ref{tab:all_coeff}, the $p$ factors are derived with values of 0.070 and 0.087 for PL1 and PLX1 centers, respectively. Though there is always a discrepancy between experimental and theoretical results, our results and analysis still contribute to the further understanding of complex physical systems when containing the JT effect. Besides, based on data displayed in Table~\ref{tab:all_coeff}, the $q$ factors are calculated and show the values of 0.507 and 0.513 for PL1 and PLX1 centers, respectively.

\begin{figure} [t]
\centering
\includegraphics[width=\linewidth]{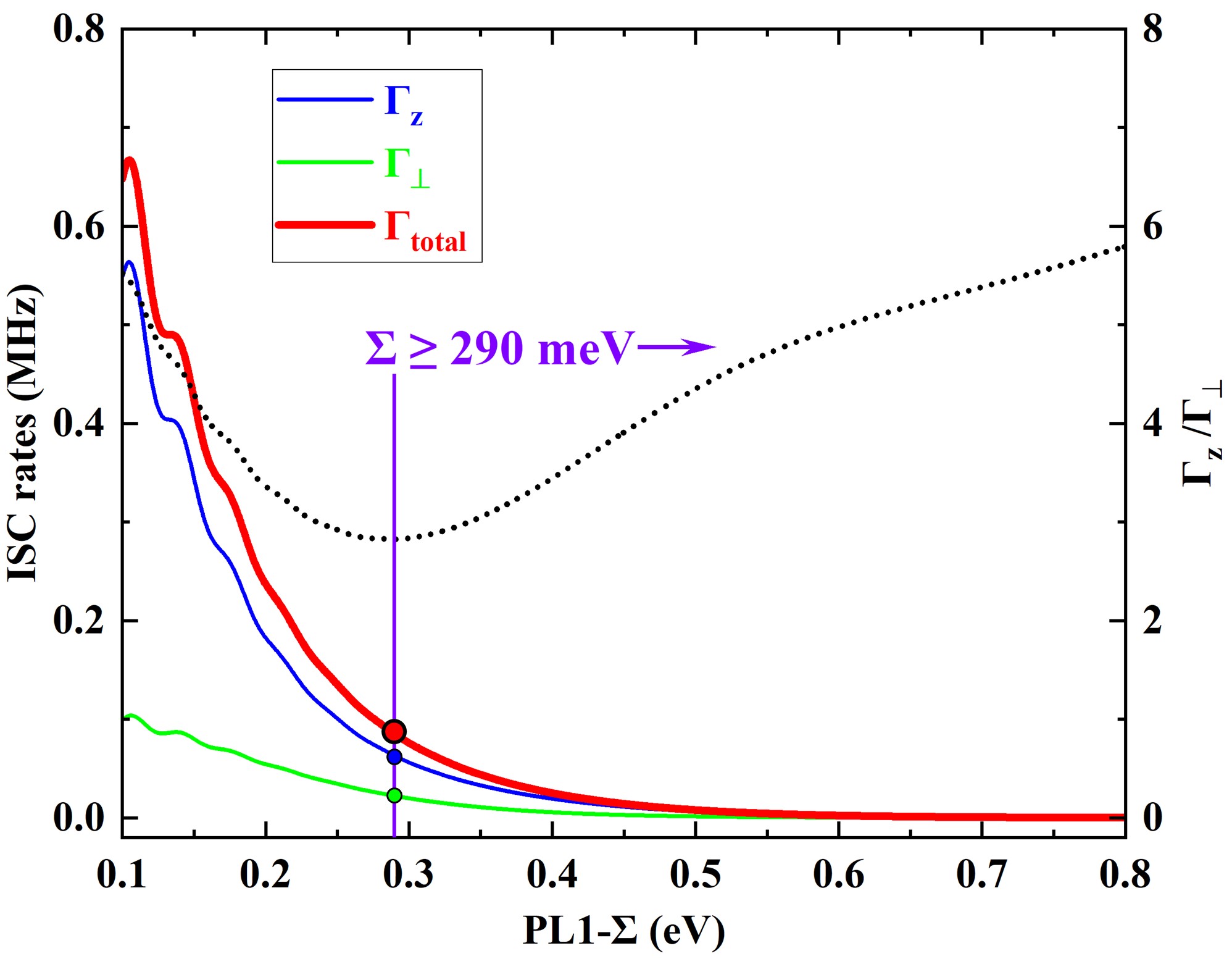}
\caption{Calculated ISC rates and $\Gamma_z / \Gamma_\bot$ ratios in the lower branch of PL1 center utilizing parameters in Ref.~\onlinecite{bockstedte2018ab}.}
\label{fig:lower_PL1_CICRA}
\end{figure}
Table~\ref{tab:all_coeff} displays the electron-phonon coupling coefficients defined in Eq.~\eqref{eq:E_pm_vibr} in the lower branch of the ISC transition. The determination of the phonon limitation is $i = 10$ to satisfy numerical convergence. The coefficient $g'_i{}^2$ for the lower branch is very small and can be ignored. Ref.~\onlinecite{bockstedte2018ab} also provides key parameters obtained from the multiconfigurational approach, including $\Sigma_{\rm{PL1'}} = 290$~meV, $\Lambda = 620$~meV and $C^2 = 0.88$. The larger $\Sigma_{\rm{PL1'}}$ in comparison to this work may arise from the fact that the JT effect is not involved and the vertical transition is considered there. Combining the vibrational calculations of this work, the lower branch rate is demonstrated and shown in Fig.~\ref{fig:lower_PL1_CICRA}. From Fig.~\ref{fig:lower_PL1_CICRA}, we find that ${\Gamma _z (\Gamma _\bot) = 0.06 (0.02)}$~MHz and ${\Gamma _z}/{\Gamma _\bot} = 2.83$ at $\Sigma_{\rm{PL1'}} = 290$~meV, consistent with the findings of this work.


\bibliography{AAA_references}

\end{document}